\documentclass[acmsmall,screen]{acmart}

\AtBeginDocument{%
  }

\setcopyright{acmlicensed}
\copyrightyear{2018}
\acmYear{2018}
\acmDOI{XXXXXXX.XXXXXXX}
\acmConference[Conference acronym 'XX]{Make sure to enter the correct
  conference title from your rights confirmation email}{June 03--05,
  2018}{Woodstock, NY}
\acmISBN{978-1-4503-XXXX-X/2018/06}

\usepackage{adjustbox}
\usepackage{amsmath,amssymb,amsfonts}
\usepackage[linesnumbered,ruled,vlined]{algorithm2e}
\usepackage{setspace}
\usepackage{algorithmic}
\usepackage{float}
\usepackage{graphicx}
\usepackage{textcomp}
\usepackage{xcolor}
\usepackage{multirow}
\usepackage{mdframed}
\usepackage{balance}
\usepackage{booktabs}
\usepackage{tcolorbox}
\usepackage{makecell}
\usepackage{threeparttable}
\usepackage{colortbl}
\usepackage{subfigure}
\usepackage{hhline}
\usepackage{pifont}
\usepackage{listings}
\usepackage{enumitem}
\usepackage{tcolorbox}
\usepackage{subcaption} 

\usepackage{booktabs} 
\usepackage{multirow}
\usepackage{array}    
\usepackage{siunitx}  

\newcommand{\revise}[1]{#1}

\newcommand{\todo}[1]{\textcolor{red}{TODO: #1}}

\definecolor{ggray}{HTML}{eff0f0}
\definecolor{gggray}{HTML}{E8E8E8}
\definecolor{ggggray}{HTML}{BEBEBE}
\definecolor{myblue}{RGB}{255,255,255}
\definecolor{myyellow}{HTML}{FFF2CC}

\newcommand{\ie}{\textit{i.e.,}}
\newcommand{\eg}{\textit{e.g.,}}

\newcommand{\claudethree}{Claude-3.5-Sonnet}
\newcommand{\claudefour}{Claude-4-Sonnet}
\newcommand{\gpt}{GPT-4.1}
\newcommand{\qwen}{Qwen3-Coder}
\newcommand{\agent}{LLM-based agent}
\newcommand{\dataset}{\textit{APPDev}}
\newcommand{\reqdoc}{requirement document}
\newcommand{\design}{overall design}

\newcommand{\ourtool}{\textsc{EvoDev}}
\newcommand{\meta}{MetaGPT}
\newcommand{\gpte}{GPT-Engineer}
\newcommand{\map}{feature map}
\newcommand{\filecontent}{\texttt{file\_contents}}

\newcommand{\improveclaudecode}{57.3}
\newcommand{\improvesingle}{58.5}

\newcommand{\redCross}{\textcolor{red}{\ding{55}}}
\newcommand{\greenCheck}{\textcolor{upgreen}{\ding{51}}}

\newcommand{\finding}[2]{
 	\vspace{1mm}
	\begin{mdframed}[linecolor=gray,roundcorner=12pt,backgroundcolor=gray!15,linewidth=3pt,innerleftmargin=2pt, leftmargin=0cm,rightmargin=0cm,topline=false,bottomline=false,rightline = false]
		\textbf{Answer to RQ#1:} #2
	\end{mdframed}
}

\lstdefinelanguage{json}{
  basicstyle=\ttfamily\small,
  numbers=left,
  numberstyle=\tiny,
  stepnumber=1,
  showstringspaces=false,
  breaklines=true,
  literate=
   *{0}{{{\color{numb}0}}}{1}
    {1}{{{\color{numb}1}}}{1}
    {2}{{{\color{numb}2}}}{1}
    {3}{{{\color{numb}3}}}{1}
    {4}{{{\color{numb}4}}}{1}
    {5}{{{\color{numb}5}}}{1}
    {6}{{{\color{numb}6}}}{1}
    {7}{{{\color{numb}7}}}{1}
    {8}{{{\color{numb}8}}}{1}
    {9}{{{\color{numb}9}}}{1}
    {:}{{{\color{punct}{:}}}}{1}
    {,}{{{\color{punct}{,}}}}{1}
    {\{}{{{\color{delim}{\{}}}}{1}
    {\}}{{{\color{delim}{\}}}}}{1}
    {[}{{{\color{delim}{[}}}}{1}
    {]}{{{\color{delim}{]}}}}{1},
}

\definecolor{numb}{rgb}{0.37,0.56,0.74}
\definecolor{punct}{rgb}{0.00,0.00,0.00}
\definecolor{delim}{rgb}{0.50,0.50,0.50}
\definecolor{string}{rgb}{0.06,0.57,0.06}

\definecolor{myorgange}{RGB}{251,229,214}
\definecolor{mygreen}{RGB}{226,240,217}
\definecolor{myblue}{RGB}{218,227,243}
\definecolor{myred}{RGB}{192,0,0}
\definecolor{mygray}{RGB}{230,230,230}
\definecolor{upgreen}{RGB}{34,139,34}

\usepackage{soul}

\usepackage{multirow}
\usepackage[normalem]{ulem}
\useunder{\uline}{\ul}{}

\newcommand{\distance}{4pt}
\begin{document}

\title{Towards Iterative End-to-End Software Development: A Feature-Driven Multi-Agent Framework}

\author{Junwei Liu}
\email{jwliu24@m.fudan.edu.cn}
\affiliation{%
  \institution{Fudan University}
  \city{Shanghai}
  \country{China}
}

\author{Chen Xu}
\email{angrytowritecode@gmail.com}
\affiliation{%
  \institution{Fudan University}
  \city{Shanghai}
  \country{China}
}


\author{Chong Wang}
\authornote{Chong Wang and Xin Peng are corresponding authors.}
\email{chong.wang@ntu.edu.sg}
\affiliation{%
  \institution{Nanyang Technological University}
  \city{Singapore}
  \country{Singapore}
}

\author{Tong Bai}
\email{22300130109@m.fudan.edu.cn}
\affiliation{%
  \institution{Fudan University}
  \city{Shanghai}
  \country{China}
}

\author{Weitong Chen}
\email{chenwt25@m.fudan.edu.cn}
\affiliation{%
  \institution{Fudan University}
  \city{Shanghai}
  \country{China}
}

\author{Kaseng Wong}
\email{alsowereme@outlook.com}
\affiliation{%
  \institution{Fudan University}
  \city{Shanghai}
  \country{China}
}

\author{Yiling Lou}
\email{yilinglou@fudan.edu.cn}
\affiliation{%
  \institution{Fudan University}
  \city{Shanghai}
  \country{China}
}

\author{Xin Peng}
\authornotemark[1]
\email{pengxin@fudan.edu.cn}
\affiliation{%
  \institution{Fudan University}
  \city{Shanghai}
  \country{China}
}

\renewcommand{\shortauthors}{Trovato et al.}

\begin{abstract}
Recent advances in large language model agents offer the promise of automating end-to-end software development from natural language requirements. However, existing approaches largely adopt linear, waterfall-style pipelines, which oversimplify the iterative nature of real-world development and struggle with complex, larger-scale projects. To address these limitations, we propose \ourtool{}, an iterative software development framework inspired by feature-driven development. \ourtool{} decomposes user requirements into a set of user-valued features and constructs a Feature Map, a directed acyclic graph that explicitly models dependencies between features. Each feature node in the feature map maintains multi-layer contexts, including business logic, software design, and code implementation, which are propagated along dependencies to provide context for subsequent development iterations. We evaluate \ourtool{} on challenging Android development tasks and show that it outperforms the best-performing baseline, Claude Code, by \revise{\improveclaudecode{}}\%, while improving single-agent performance by 16.0\%–\revise{\improvesingle{}}\% across different base LLMs, highlighting the importance of feature decomposition, dependency modeling, context propagation, and workflow-aware agent design for end-to-end software development. Moreover, our work summarizes practical insights for designing iterative, LLM-driven development frameworks and informs future training of base LLMs to better support iterative software development. 
\end{abstract}



\received{20 February 2007}
\received[revised]{12 March 2009}
\received[accepted]{5 June 2009}

\maketitle

\section{Introduction}
Software development has long been regarded as a complex, resource-intensive task that relies heavily on collaboration among domain experts~\cite{whitehead2007collaboration,bialy2017software,evans2004domain}, such as software architects, programmers, and testers.
Recent advances in large language model (LLM) agents have opened up new opportunities for automatic \textit{end-to-end software development} tasks, which aim to produce executable software directly from natural language requirements~\cite{agent4se}. By simulating real-world development processes, multi-agent workflows can translate user requirements into executable applications.

However, most prior approaches follow a linear, waterfall-style pipeline in which requirements analysis, design, implementation, and testing are executed sequentially~\cite{MetaGPT, ChatDev, AISD, Claude-Code, liualtdev}.  This workflow oversimplifies the dynamic and iterative nature of real-world software development and fails to handle the development of more \revise{realistic} software.
Some research explores the possibility of designing an agile workflow to conduct iterative software development. For example, AgileCoder~\cite{AgileCoder} designs a Scrum-inspired workflow, where the software development proceeds in iterative sprints. However, it primarily relies on the capability of the underlying model to identify relevant context and extension points within the increasingly larger code repository, which results in its limited capability in developing \revise{ applications with more features and constraints.}

Moreover, existing works still focus on relatively simple development scenarios, such as common and small-scale Python applications or web pages (\eg{} snake game), which fall short of the complexity encountered in real-world software development~\cite{agent4se,peng2025code}. For instance, MetaGPT~\cite{MetaGPT} has been applied to develop Python applications with fewer than 250 lines of code. In contrast, more complex development scenarios, such as Android development, pose additional challenges. Implementing features on Android often requires handling intricate application lifecycles, coordinating asynchronous tasks, managing diverse dependency configurations, and integrating with platform-specific APIs. However, the effectiveness of \agent{}s in such more complex development scenarios remains largely unexplored.

To address these challenges, we propose \ourtool{}, an iterative software development framework inspired by the classic feature-driven development (FDD) methodology~\cite{goyal2007agile}. Specifically, we first decompose the original user requirements into user-valued features and then construct a directed acyclic graph (DAG) called \textbf{Feature Map} to explicitly model the dependencies between these features. Each node in the feature map stores multi-level information about a feature, including business logic, design, and code implementation. Through the dependency relationships, this information \revise{propagates} to subsequent features, providing the necessary context to support smooth iterative development.

To evaluate the effectiveness and efficiency of \ourtool{}, we conduct experiments on more challenging Android development tasks, \revise{targeting the development of complete client applications with more sophisticated business functionality.} 
Experimental results demonstrate that \ourtool{} outperforms all existing LLM-agent baselines, and surpasses the best-performing baseline, Claude Code, by \revise{\improveclaudecode}\%. In addition, \ourtool{} brings consistent improvements to the single agent baselines by a relative improvement ranging from 16.0\% to \revise{\improvesingle}\% across different base LLMs. We also summarize practical implications based on our experimental results to guide the design of an LLM-driven iterative development framework and the training of base LLMs.

In summary, this work makes the following contributions:
\begin{itemize}
    \item We propose and implement the first FDD-inspired iterative software development framework, \ourtool{}, which constructs a global feature map to store and propagate contexts of different abstract layers through dependencies between iterations.
    \item We manually construct an Android software development dataset, \dataset{}, and conduct a comprehensive set of experiments to evaluate the effectiveness and efficiency of \ourtool{}. 
    \item We summarize key implications of our experimental findings for future practice and research, including the demands that iterative development places on model capabilities and context management, the misalignment of model-intrinsic behaviors with workflow guidance, and the realities of iterative development cost.
\end{itemize}

\section{Background and Related Work}
\subsection{LLM-based agents for End-to-end Software Development}
\agent{}s have emerged as a promising solution for software engineering tasks~\cite{agent4se}. With enhanced perception, planning, memory, and action, these agents can act as virtual programmers, supporting diverse development activities such as requirements engineering~\cite{abedini2025leveraging,tauhid2025explainability}, coding~\cite{DBLP:journals/corr/abs-2406-10018,ciniselli2024generalizability,da2025llms}, testing~\cite{baudry2024generative,abdullin2025test}, and debugging~\cite{haryono2021androevolve,lin2025codereviewqa,sejfia2024toward}.
Building on this progress, recent research has begun to explore their capabilities in
end-to-end software development, which focuses on generating target software based on user requirements from scratch~\cite{agent4se}.
To cover diverse activities in the software development process, most existing studies emulate the real-world development process and design multi-agent workflows. Among these workflows, the linear workflow, which is inspired by the classic Waterfall process model~\cite{petersen2009waterfall}, is most frequently used. It decomposes the entire development process into sequential stages such as requirements analysis, design, coding, and testing, and assigns one or more agents to each stage. Representative work includes MetaGPT~\cite{MetaGPT} and ChatDev~\cite{ChatDev}. 
In addition, some general-purpose software engineering agents like GPT-Engineer~\cite{GPT-Engineer} and Claude Code~\cite{Claude-Code}, also adopt the linear workflow to ensure versatility. For example, Claude Code always outputs a linear to-do list first and completes the task step by step.
However, this linear workflow oversimplifies the dynamic and iterative nature of real-world software development and faces challenges in larger-scale software projects. In addition, it relies heavily on the capability of base LLMs to produce a comprehensive code repository covering all aspects of the user requirements.

Another research direction is to integrate an agile workflow with \agent{}s.
For example, AgileCoder~\cite{AgileCoder} designs a Scrum-inspired iterative workflow, with each sprint containing planning, development, testing, and review.
\revise{However, AgileCoder lacks effective context management and mainly relies on the underlying model to locate modification points in the code repository, which has been proven to be a challenging task~\cite{lingma, rahardja2025can}. 
As a result, AgileCoder can merely handle limited software complexity, with most examples still being common and simple applications (\eg{} snake game), which is far from supporting truly iterative and incremental development.
}
\revise{Another attempt is RPG~\cite{luo2025rpg}, which models the target software repository as a tree and extends data flows and interfaces to support iterative development via topological ordering. However, the tree structure favors modular decomposition, making it suitable mainly for tool libraries with clear module boundaries (\eg{} pandas, scikit-learn) rather than client applications with complex business logic.
Therefore, to align with real-world iterative software development and adapt to more complex client-valued software, it is important to implement a practical iterative development framework. 
}

\subsection{Feature-Driven Development}
\begin{figure}[htb]
    \centering
    \includegraphics[width=0.9\columnwidth]{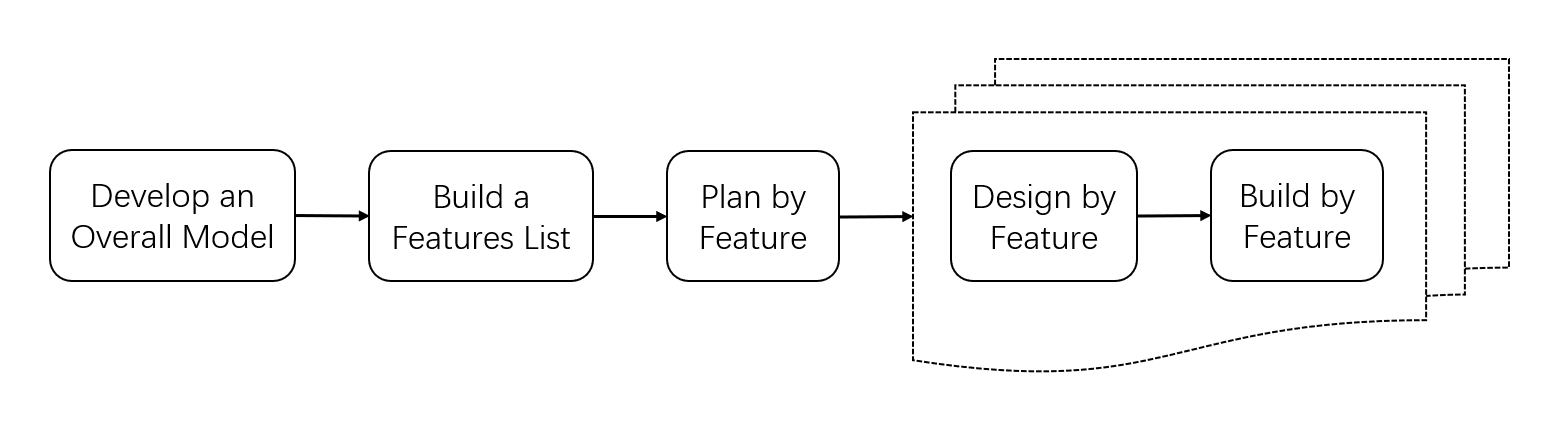}
    \vspace{-3mm}
    \caption{The basic activities of Feature Driven Development} 
    \label{fig:FDD_background}
\end{figure}

Feature-driven development (FDD) is an agile software development methodology~\cite{goyal2007agile}.
As shown in Figure~\ref{fig:FDD_background}, the FDD workflow starts with constructing an overall model to analyze the business relationships and establish a consistent design. After that, a list of \textit{features} is extracted, with each feature representing a small user-valued functionality. Functionally cohesive features can be further aggregated into feature sets. Next, team members carefully consider the dependencies and priorities of the features and plan their development order. Finally, in each iteration, the development team makes a fine-grained design of the current feature and builds it.

Despite a classic methodology for human teams, FDD is well-suited for adaptation to a multi-agent development workflow since the input and output of all activities can be formulated in pure text format to be processed by \agent{}s. 
In addition, FDD presents a promising paradigm for \agent{}s to tackle the complexities of end-to-end software development. First, all of its activities are grounded in the initial overall design, which helps maintain consistency and alignment throughout the iterative development cycles. Second, it breaks down complex requirements into small, manageable, and verifiable features, reducing the implementation difficulty in each iteration. Third, it includes a dedicated activity to model feature dependencies and determine development order (\ie{} plan by feature), which helps maintain a smooth and reliable development process. These dependencies can further help agents to retrieve necessary contexts from preceding features, which often include the interface for implementing the current feature. Finally, it balances the global overall design with the fine-grained design within each iteration, alleviating the burden and difficulty of producing a detailed upfront design. 
Therefore, inspired by the strengths and adaptation, we design a multi-agent framework to simulate the FDD workflow.

\section{FDD-Inspired Iterative Development Framework} \label{sec:method}
\begin{figure}[htb]
    \centering
    \includegraphics[width=0.9\columnwidth]{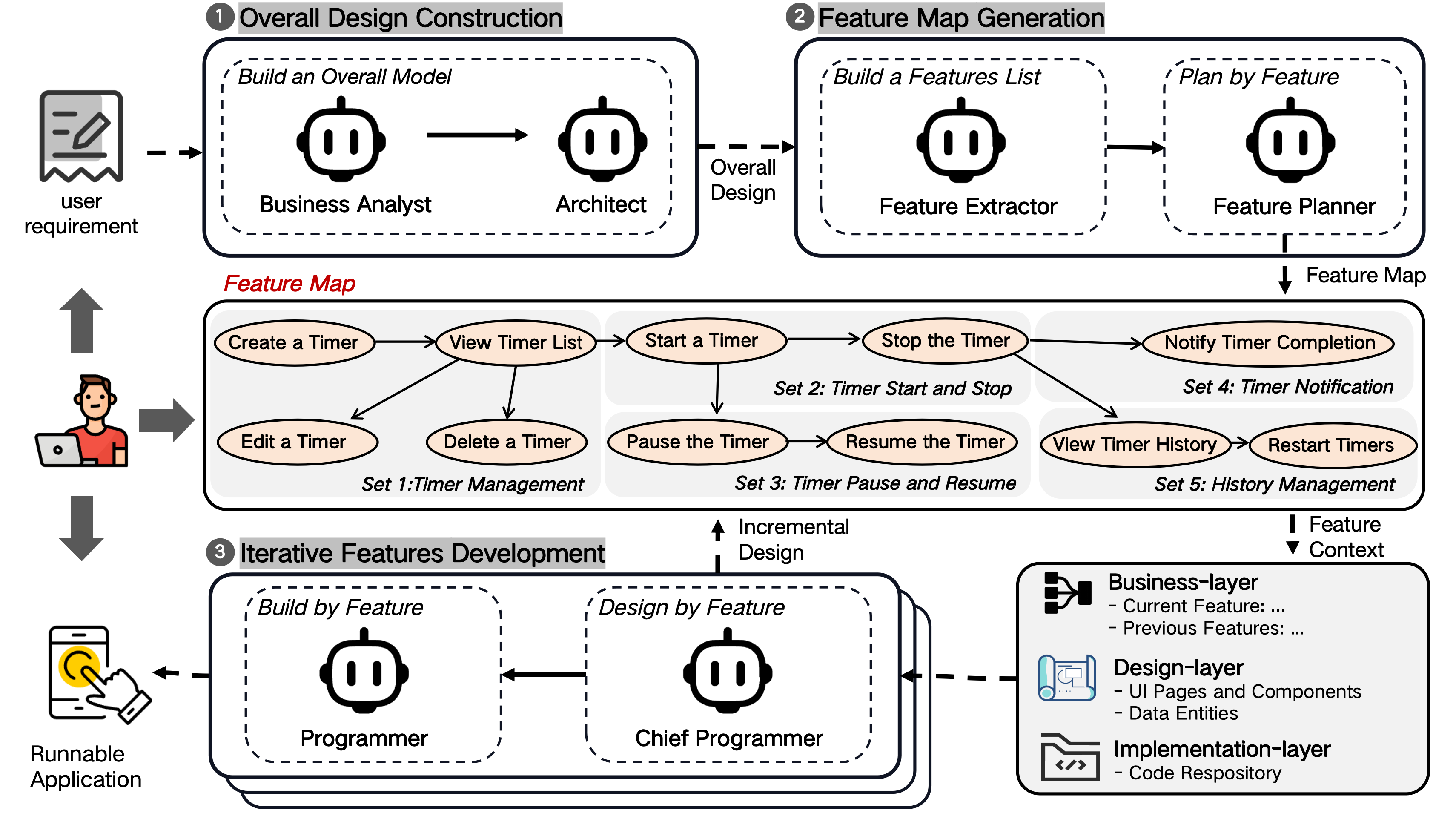}
    \caption{Overview of the FDD-inspired \ourtool{} framework} 
    \label{fig:overview}
\end{figure}

In this section, we present \ourtool{}, an FDD-inspired iterative software development framework. As illustrated in Figure~\ref{fig:overview}, \ourtool{} abstracts the original FDD process into three stages: \textbf{Overall Design Construction}, \textbf{Feature Map Generation}, and \textbf{Iterative Features Development}. Each stage corresponds to one or two steps in the standard FDD workflow, as indicated by dashed boxes.

\vspace{3mm}
\vspace{-3mm}

\subsection{Overall Design Construction}\label{sec:method:overall_design} 
\begin{figure}[htb]
    \centering
    \begin{minipage}{0.45\textwidth}
        \centering
        \includegraphics[width=\linewidth]{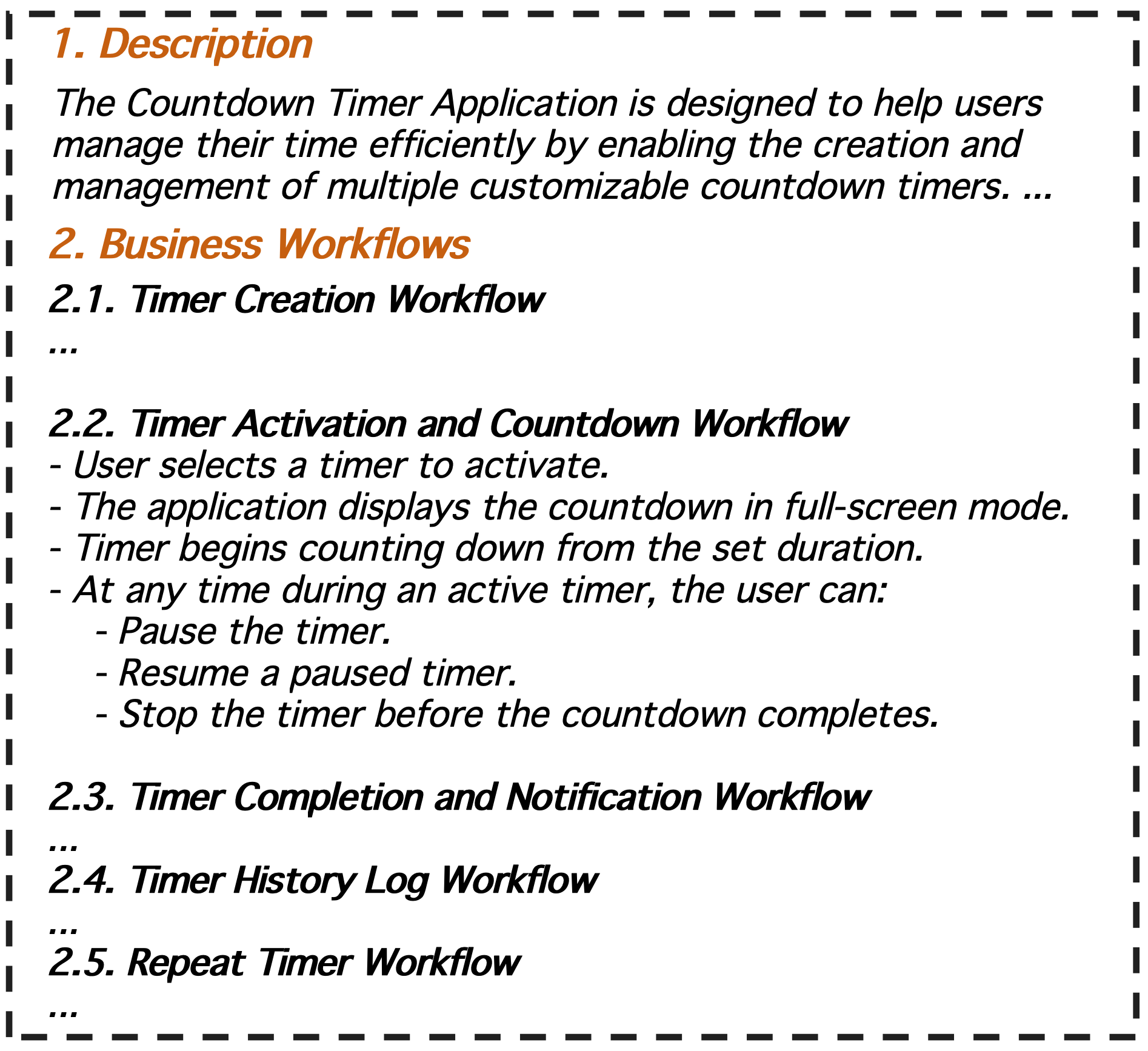}
        \caption{The requirement document for the \textit{countdown timer} APP}
        \label{fig:requirement_document}
    \end{minipage}%
    \hspace{0.01\textwidth}
    \begin{minipage}{0.45\textwidth}
        \centering
        \includegraphics[width=\linewidth]{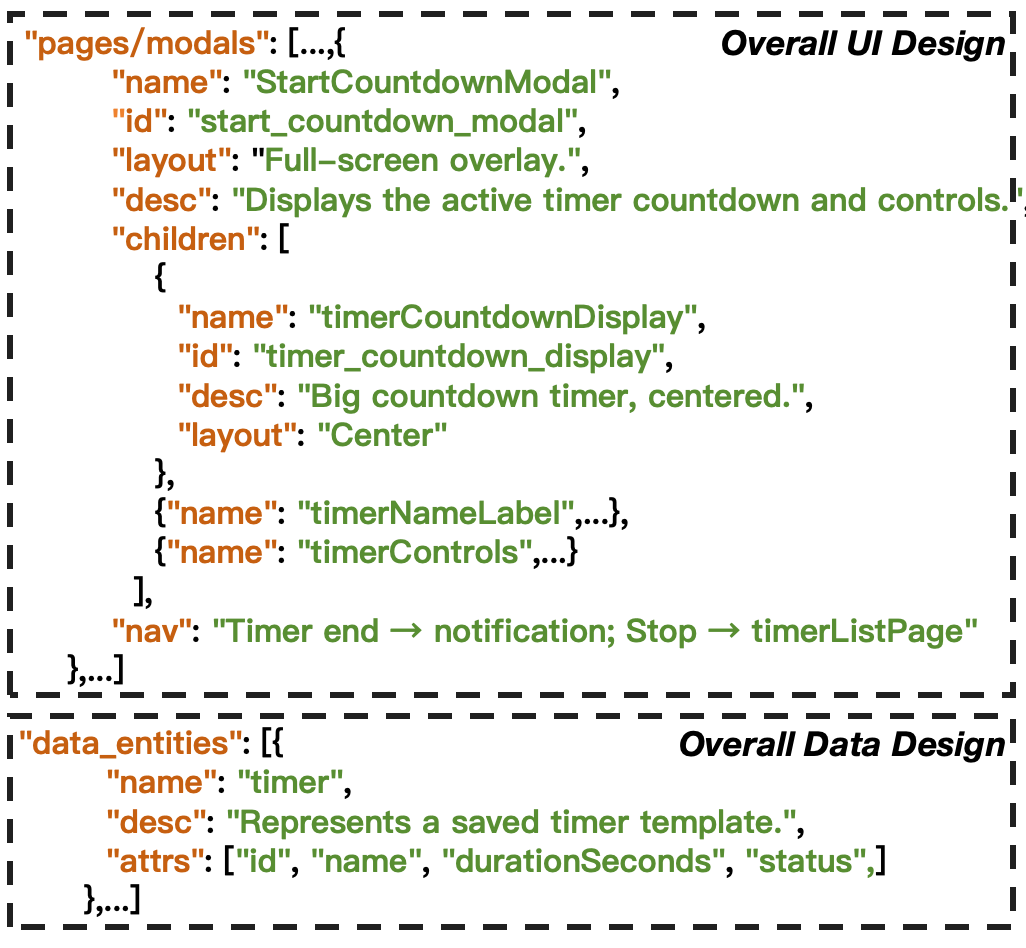}
        \caption{The overall design (including both UI and data) for the \textit{countdown timer} APP}
        \label{fig:overall_design}
    \end{minipage}
\end{figure}

The first stage of \ourtool{} is to construct an overall design of the target application based on the user requirements, which corresponds to the ``\textit{Build an Overall Model}'' step in FDD.
However, the user-provided requirements are often informal, with information scattered and loosely connected. To address this, we add a \textbf{Business Analyst} agent to analyze the user-provided requirements and generate a structured \reqdoc{}. As illustrated in Figure~\ref{fig:requirement_document}, the \reqdoc{} consists of two parts: (i) a concise description of the target application and (ii) a comprehensive list of all identified business workflows. Through this step, we reorganize the user-provided informal input into a structured document, facilitating subsequent data processing. 

  

Next, we introduce an \textbf{Architect} agent to construct an overall design of the target application.
In FDD, the overall model refers to the domain object model. However, our preliminary experiments indicate that LLMs struggle to generate a usable domain object model, which is also reported in prior work~\cite{LLM-Based_Domain_Modeling,ge2023openagi,chen2023automated}. Therefore, we shift to leveraging LLMs for data and user interface (UI) design.
A key challenge in software development is that capturing all design details in the initial stage is non-trivial. To address this, we instruct the Architect agent to generate an initial \design{} of the target application. As illustrated in Figure~\ref{fig:overall_design}, the \design{} includes two parts: (i) a coarse-grained UI design with only the essential pages and the top-tier components on each page, and (ii) the necessary data entities.
This \design{} serves two primary purposes. First, it acts as the blueprint for subsequent development, preventing agents from misinterpreting the relationships between data models and UI components and arbitrarily altering them in different iterations. Second, it establishes a shared communication vocabulary among subsequent agents, which helps maintain consistency during the iterative development process.

\subsection{Feature Map Generation}\label{sec:method:feature_map}
Based on the constructed overall design, \ourtool{} continues to generate a global feature dependency graph, \textbf{Feature Map}, to guide the iterative development, corresponding to the ``\textit{Build a Features List}'' and ``\textit{Plan by Feature}'' steps in the FDD workflow.

\begin{figure}[htb]
    \centering
    \includegraphics[width=0.8\columnwidth]{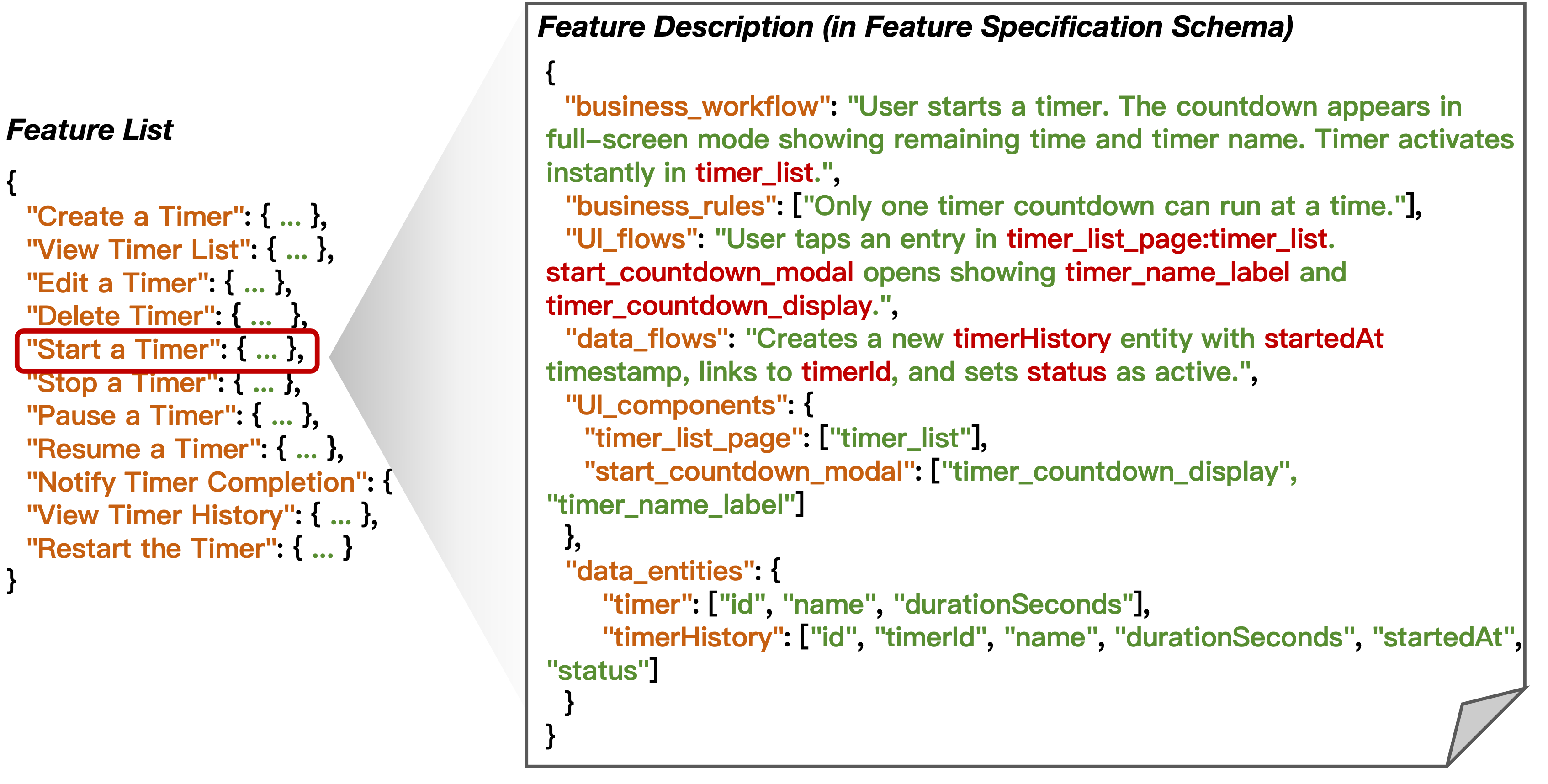}
    \caption{The feature list for the countdown timer APP. The \textbf{\textcolor{myred}{bold red}} highlights indicate the UI and data designs referenced in the UI flow and data flow descriptions.} 
    \label{fig:feature_list}
\end{figure}

First, we employ a \textbf{Feature Extractor} agent to extract the feature list. This agent decomposes the original requirements and overall design into a set of features, with each feature representing a small, client-valued functionality. To clarify the functional scope of each feature, we design a \textit{Feature Specification Schema}, which includes the following fields:
\begin{itemize}[leftmargin=*]
    \item \textit{Business workflow}: Describes the business workflow and the final delivery target.
     \item \textit{Business rules}: Highlights rules that capture the details of the current feature, such as the default values for user data and validation rules.
    \item \textit{UI flow}: Details how the user interacts with the UI components defined in the \design{}.
    \item \textit{Data flow}: Details how the data entities in the \design{} are processed within this feature.
    \item \textit{Contained data models and UI components}: Enumerates IDs of all data models and UI components involved in the feature, which are used to extract the corresponding contexts in \design{}.
\end{itemize}
To ensure consistency, all descriptions reference the IDs of UI components and data models from the overall design. Figure~\ref{fig:feature_list} shows the feature list of the \textit{Countdown Timer} application and the description of the ``\textit{Start a Timer}'' feature, with bold red text indicating the corresponding IDs.

\begin{figure}[htb]
    \centering
    \includegraphics[width=0.8\columnwidth]{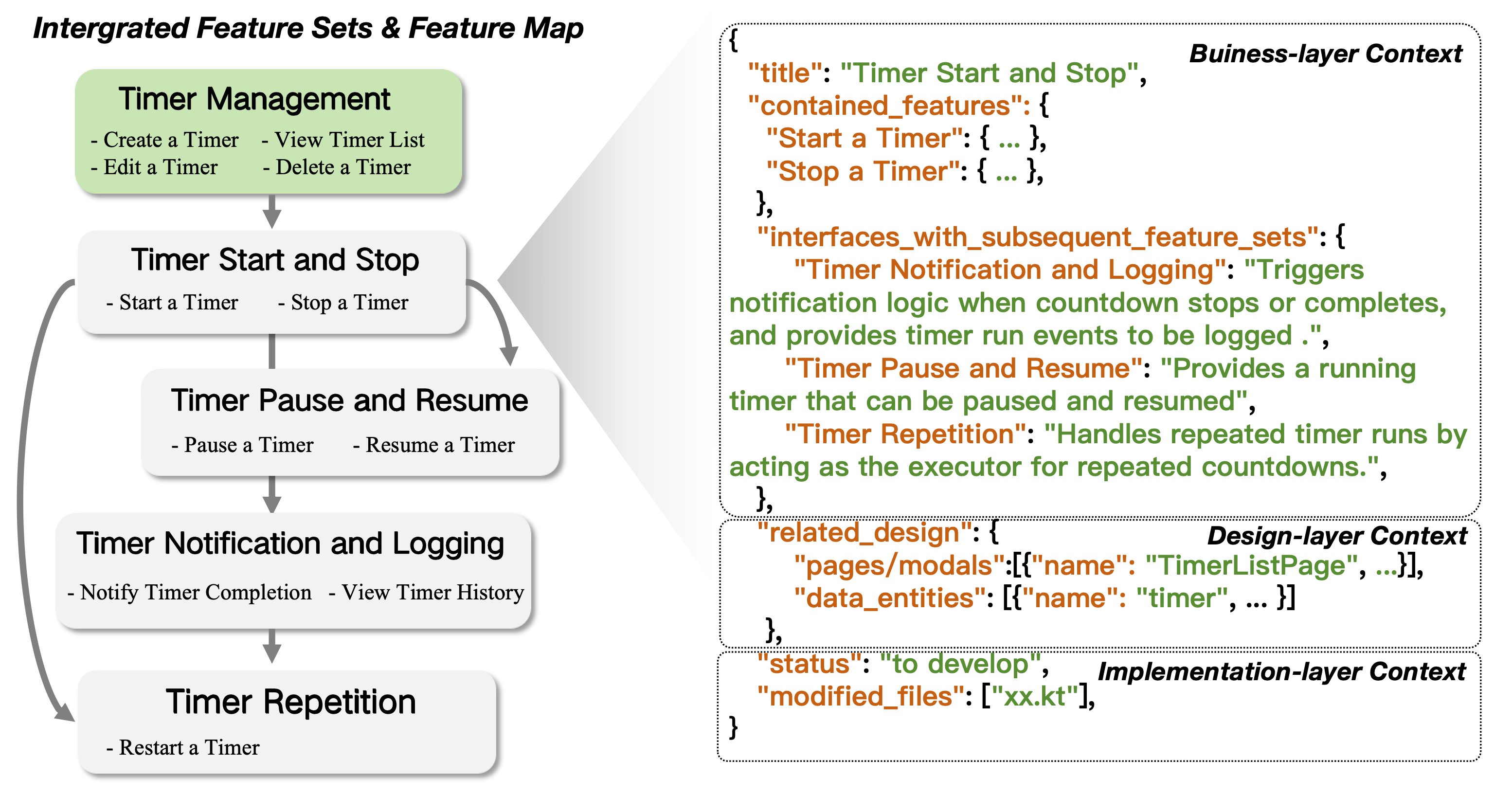}
    \caption{The feature map for the countdown timer APP, with each node containing contexts of the business, design, and implementation layers.} 
    \label{fig:feature_set}
\end{figure}

Afterwards, we assign a \textbf{Feature Planner} agent to organize the development of the extracted features. 
First, the Feature Planner analyzes dependencies among extracted features from both \textit{business} and \textit{technical} perspectives, as illustrated in the central section of Figure~\ref{fig:overview}.
Second, it adopts a bottom-up approach to group these features into cohesive feature sets and determines their dependencies by following two principles: (i) features within a set should exhibit high cohesion, allowing them to be implemented within a single iteration; (ii) feature sets must preserve the original feature-level dependencies. For example, if Feature-B depends on Feature-A, then Feature-A must either reside in the same set as Feature-B or in a preceding set.  
One advantage of this approach is that the size of each feature set can be tailored to the coding capacity of different LLMs, which enhances the overall adaptability and versatility of this framework.
Finally, based on the dependencies between feature sets, our framework generates a directed acyclic graph (DAG) over the extracted feature sets, which is referred to as a \textbf{Feature Map}.
As illustrated in Figure~\ref{fig:feature_set}, each feature set in the feature map includes three different layers of context:
\begin{itemize}[leftmargin=*]
    \item \textbf{Business-layer Context}: this context includes the constituent features of this set and its interfaces with subsequent feature sets, specifying the functional scope of this set.
    \item \textbf{Design-layer Context}: this context extracts all UI components and data entities involved in constituent features from the global \design{}.
    \item \textbf{Implementation-layer Context}: this context includes the development status and the modified files of the current feature, which helps locate relevant code for subsequent features.
\end{itemize}

In contrast to the widely adopted linear planning strategies that merely produce a to-do list, our \map{} not only captures the complex dependencies among requirements using a graph structure, but also encapsulates critical context at the business, design, and implementation layers. This enables agents to track repository changes across both temporal (feature dependencies) and spatial (design and code changes) dimensions, thereby reducing the effort required for the agents to understand the complex code repository and improving development accuracy.

\subsection{Iterative Features Development}\label{sec:method:feature_developement}
The final stage of our framework is to iteratively develop the target application. Following the FDD methodology, where each iteration consists of ``\textit{Design by Feature}'' and ``\textit{Build by Feature}'' steps, we first perform a detailed design of the current feature set, and subsequently implement it based on the design specifications.


We assign a \textbf{Chief Programmer} agent to make a fine-grained design of the current feature set, which takes the business-layer and design-layer context of both the current set and its preceding sets as input and completes three tasks. First, it adopts a bottom-up method to analyze and integrate all features within the current feature set and output a feature-set-level description following the \textit{Feature Description Schema} introduced in Section~\ref{sec:method:feature_map}.
Second, it generates incremental design descriptions for the involved components (\eg{} “\textit{Add a rounded confirm button with a green background at the bottom of the original modal.}”), which are then incorporated back into the \design{} and available to subsequent iterations. This ensures that the overall design provides global guidance while allowing the Chief Programmer agent to make flexible, iteration-specific design decisions.
Finally, the Chief Programmer further decomposes the fine-grained design into a development plan consisting of all tasks to complete in this iteration.
In summary, the Chief Programmer reorganizes the relationships among all features in the current feature set, extends the relevant coarse-grained overall design, and produces a detailed development plan for the current iteration.

We then design a \textbf{Programmer} agent to take over responsibility for the concrete implementation. The Programmer takes the fine-grained design and development plan produced by the Chief Programmer, along with the implementation-layer context (\eg{} changes in the code repository) as input, and is equipped with tools for reading, writing, and editing code files.
To further improve efficiency, we design a memory mechanism to optimize the agent's trajectory management.
In particular, we observe that when an agent invokes tools, its response includes a \texttt{tool\_calls} field that encompasses all the tool names and parameters. This field stores fragmented code modifications, which quickly bloat the context, introduce redundant file versions, and increase reasoning complexity. To address this, we maintain a unique memory called \filecontent{} to cache the latest version of all files read, created, or edited by the agent. When the agent invokes tools, the \texttt{tool\_calls} will be executed and removed, leaving merely the natural language response (e.g., “\textit{Let’s modify [FILE\_PATH] to implement the [FUNCTION]}”) in the dialog history. We then insert an extra message to report the tool execution result and notify that \filecontent{} has been updated. This ensures trajectory completeness while keeping unique, always up-to-date file versions in the dialogue history, thereby maintaining a clear context with fewer tokens.
Equipped with the memory, the Programmer implements the current feature sets through two phases: coding and debugging. During the coding stage, the Programmer autonomously invokes tools to modify code in the repository. Upon completion, the debugging phase is triggered. The developed application is automatically built, with the error messages returned as feedback for the Programmer to revise the implementation, after which the build is automatically re-run. This loop continues until all errors are resolved. 

Finally, \ourtool{} performs a git commit, updates the development status and modified files of the current feature set, and delivers a runnable application that can be experienced by the user. The iterative development process continues until all feature sets in the feature map have been implemented and the final application is delivered.

\section{Experimental Setup}
In this section, we describe the setup of a comprehensive set of experiments to demonstrate the effectiveness and efficiency of our proposed \ourtool{} framework in iterative software development. 

\subsection{Research Question}
We formulate the following research questions to guide our experiments:

\begin{itemize}[leftmargin=*]
    \item \textbf{RQ1-Effectiveness}: How effective is our FDD-inspired framework, \ourtool{}, compared with other LLM-agent-based approaches for software development?

    \item \textbf{RQ2-Generalizability}: Does \ourtool{} demonstrate consistent effectiveness in different settings?
        \begin{itemize}
            \item \textbf{RQ2.a-Model Generalizability}: Does \ourtool{} demonstrate consistent effectiveness in software development across different base LLMs?
            \item \textbf{RQ2.b-Task Generalizability}: Does \ourtool{} demonstrate consistent effectiveness when applied to tasks of different difficulties?
        \end{itemize}

    \item \textbf{RQ3-Ablation Study}: How do different stages impact the performance of \ourtool{}?
    
    \item \textbf{RQ4-Efficiency}: How efficient is \ourtool{}, as an iterative approach, compared with other single-pass approaches for software development?

    \item \revise{\textbf{RQ5-Failure Analysis}: How do different agents differ in development failure modes?}
\end{itemize}

\subsection{Implementation Details}
\textbf{Software Platform Selection.} 
\revise{We choose Android development as our test scenario, primarily for three reasons.}
First, Android powers billions of devices worldwide, making it one of the most widely adopted software platforms~\cite{AbdulKadir2024,han2024survey}. However, it has never been evaluated in any end-to-end software development research. 
Second, compared to widely evaluated Python and Web development, Android development is inherently complex, involving strict system constraints, diverse hardware interactions, and intricate application lifecycles. 
Third, the default programming language for Android development is Kotlin, which is a low-resource programming language with limited corpora available for training~\cite{Kotlin-ML}.
These factors together make Android a more challenging testbed to evaluate the effectiveness and generality of \agent{}s in software development.

\textbf{Workflow Implementation.} We implement \ourtool{} for Android development on top of \textit{LangGraph}~\cite{langchain_langgraph_2024}, which is widely used in prior work~\cite{mandulapalli2025development,sharma2025optimised,akilesh2025multi}. 
The workflow requires two inputs: (i) a natural language requirements description; and (ii) the path to a scaffold project generated by the official Android Studio IDE, serving as the execution environment. 
Except for the requirement document, all intermediate outputs are stored in JSON output, which facilitates management and context retrieval.
\revise{Based on preliminary experiments, we limit the number of extracted feature sets to no more than four to balance effectiveness and efficiency.}
For tool integration, we implement tools to read, create, and edit files. Following Claude Code~\cite{Claude-Code}, we adopt a search–substitute strategy to apply code diffs, where the agent outputs both original and revised code blocks for automatic substitution.
 
\subsection{Evaluation Benchmark}
\begin{table}[t]
\centering
\caption{Comparison of existing datasets and \dataset.}
\label{tab:benchmark}

\begin{adjustbox}{width=1.0\textwidth}
\revise{
\begin{tabular}{lccrrrrl}
\toprule
\textbf{Dataset} & \begin{tabular}[c]{@{}l@{}}\textbf{Open}\\ \textbf{Source?} \end{tabular} & \textbf{Platform} & \begin{tabular}[c]{@{}l@{}}\textbf{\#Tasks}\end{tabular} & \begin{tabular}[c]{@{}l@{}}\textbf{Avg.}\\ \textbf{Req. Len.} \end{tabular} & \begin{tabular}[c]{@{}l@{}}\textbf{Avg.}\\ \textbf{\#Features}\end{tabular} & \begin{tabular}[c]{@{}l@{}}\textbf{Avg.}\\ \textbf{\#LoC} \end{tabular} & \textbf{Metrics} \\ \midrule
\textbf{SoftwareDev~\cite{MetaGPT}} & \redCross & Python & 70 & 30 & \redCross & 251 & Program-level quality score (1–4). \\
\textbf{SRDD~\cite{ChatDev}} & \greenCheck & Python & 1200 & 55 & \redCross & 144 & Cosine similarity between code and requirements. \\
\textbf{ProjectDev~\cite{AgileCoder}} & \greenCheck & Python  & 14 & 31 & 8.9 & \redCross & Feature completion ratio (0-100\%). \\ 
\midrule
\textbf{APPDev} & \greenCheck & \textbf{Android} & \textbf{15} & \textbf{172} & \textbf{13.5} & \textbf{3656} & \textbf{Feature-level quality score (1–4).} \\ 
\bottomrule
\end{tabular}
}
\end{adjustbox}
\end{table}
\subsubsection{Dataset}
\revise{Table~\ref{tab:benchmark} summarizes the datasets used in prior work~\cite{ChatDev, MetaGPT, AgileCoder}. Through careful inspection, we find that these datasets all exhibit practical quality issues. For example, SoftwareDev and SRDD lack feature decomposition and rely on vague holistic impression or inaccurate code-requirement similarity for evaluation. ProjectDev provides overly brief task descriptions (31 words on average), with 36\% tasks consisting of merely a single sentence (\eg{} ``Create a 2048 game''), making it hard to attribute missing features to agent capability rather than insufficient requirements. In addition, these benchmarks are all built on widely trained Python programs, leaving challenging low-resourced software domains (\eg{} Android app development) unexplored.}
Motivated by these factors, we manually construct an Android application development dataset, \dataset{}, through a three-stage pipeline:

\textbf{Stage I: Application Selection.} 
After assessing the manual efforts required for evaluation, we first fixed the dataset size to 15 Android applications.
\revise{We begin with filtering applications from real-world app stores, guided by diverse criteria including \textit{category}, \textit{popularity}, \textit{user scenarios}, and \textit{development patterns}. 
Specifically, we first examined application categories in the Google Play Store and Apple App Store, identifying the five most common: Utility, Lifestyle, Shopping, Education, and Entertainment.
Within each category, applications were ranked by popularity. Candidates were selected in descending order according to the following criteria: (i) we retained only lightweight applications whose core functionalities can be implemented on the client side, without relying on backend systems or online services; (ii) instead of directly reproducing commercial applications, we distilled representative usage scenarios (\eg{} music playback) and ensured diversity in development patterns, such as third-library usage (\textit{QRCode Tool}), animations (\textit{Decision Wheel}), and concurrency (\textit{Sheet'em Up}). This stage concluded once each category contained three applications.
}

\textbf{Stage II: Requirement Specification.}
In the second stage, all authors conducted a collaborative brainstorming session to analyze the functional requirements of the selected applications, identify the core functionalities to retain, and draft an initial version of the requirement descriptions.
After that, the first two authors independently drafted core acceptance checklists for each application and reconciled their drafts through discussion. When consensus was not reached, a third author acted as an arbitrator. This iterative process continued until all three authors agreed on the final checklists.
Simultaneously, the requirement descriptions were refined in accordance with updates to the checklist, ensuring that all functional requirements were clearly conveyed.

\textbf{Stage III: Industrial Validation.}
To further ensure dataset quality, we invited professional practitioners from the industry to review the dataset, \revise{who verified that these applications can better align with real-world development practices compared to existing datasets.}

Table~\ref{tab:dataset} summarizes the constructed dataset. For each application, we provide a detailed requirement description and an acceptance checklist. Applications have an average of 13.5 functional requirements (ranging from 8 to 26) and are categorized into three difficulty levels: \textit{Elementary} (0–9), \textit{Intermediate} (10–19), and \textit{Advanced} (20+). This categorization enables evaluation of \agent{}s’ performance across applications of varying complexity.

\begin{table}[t]
\centering
\caption{Statistics of \dataset{}. ``\# Req.'' represents the number of functional requirements.}
\label{tab:dataset}
\begin{adjustbox}{width=1.0\textwidth}
\footnotesize
\begin{tabular}{>{\raggedright\arraybackslash}p{2.5cm}|l|>{\centering\arraybackslash}p{1cm}|>{\raggedright\arraybackslash}p{7.4cm}|l}
\toprule
\textbf{APP Name }             & \textbf{Difficulty} & \multicolumn{1}{l|}{\textbf{\# Req.}} & \textbf{APP Keywords} & \textbf{Type}          \\ \midrule
Color Picker          & Elementary     & 8  & Image import, magnifier dragging, color extraction and saving. & Utility       \\ \hline
QRCode Tool           & Elementary     & 8  & QR code scan and generation, URL browsing. & Utility       \\ \hline
Expense Tracker       & Elementary     & 8  & Expense logging, budget setting, visual reports. & Shopping      \\ \hline
Shopping List         & Elementary     & 8  & Shopping list creation, price tracking, list sharing. & Shopping      \\ \hline
Countdown Timer       & Elementary     & 9  & Custom timers, personal time management, notifications. & Education     \\ \hline
Decision Wheel        & Elementary     & 9  & Customized decision wheels, animations, random outcomes. & Utility       \\ \hline
PDF Reader            & Elementary     & 9  & PDF import, file grouping, document reading, zooming. & Education     \\ \hline
Music Player          & Intermediate   & 12 & Music import and management, playback control, play modes. & Entertainment \\ \hline
Trip Itinerary        & Intermediate   & 13 & Travel itinerary planning, scheduling, notifications. & Lifestyle     \\ \hline
Drink Order           & Intermediate   & 15 & Drink browsing, order customization, tokens and payment. & Shopping      \\ \hline
Diary Journal         & Intermediate   & 16 & Diary writing, rich text formatting, images, daily reminders. & Lifestyle     \\ \hline
Fitness Record        & Intermediate   & 16 & Workout planning, record tracking, statistics visualization. & Lifestyle     \\ \hline
Simulated Planting    & Advanced       & 20 & Crop planting, fields unlocking, soil management, trading system. & Entertainment \\ \hline
Simple Duolingo       & Advanced       & 25 & Language learning, lessons unlocking, streak tracking. & Education     \\ \hline
Shoot ’em Up          & Advanced       & 26 & Enemy shooting, character control, ability upgrades, game store. & Entertainment \\ \bottomrule
\end{tabular}
\end{adjustbox}
\end{table}

\subsubsection{Baselines}
We include baselines for our evaluation according to the following criteria: (i) the tool should be based on an LLM-agent architecture; (ii) it should support Android application development, excluding tools that only support Python or Web development, such as ChatDev~\cite{ChatDev} and Lovable~\cite{lovable}; (iii) it should support incremental development on an existing code repository (\ie{} the initial Android scaffold project); and (iv) the selection should include both open-source and closed-source tools. Based on these criteria, we select three representative software development \agent{}s as our baselines, as follows:

\begin{itemize}[leftmargin=*]
    \item \textbf{MetaGPT~\cite{MetaGPT}}: An open-source multi-agent framework designed for autonomous software development. It designs a waterfall-style development workflow with five stages (\ie{} analysis, design, planning, coding, and testing).  
    
    \item \textbf{GPT-Engineer~\cite{GPT-Engineer}}: An open-source coding agent that supports incremental modification on existing repositories based on the requirement descriptions provided in prompt files.
    
    \item \textbf{Claude Code~\cite{Claude-Code}}: A state-of-the-art commercial agentic coding tool released by Anthropic, which is exclusively powered by \claudefour{}. It allows users to describe desired features in natural language and can automatically generate a to-do list, write code, and verify correctness.

\end{itemize}
All of the baselines have also been frequently compared in prior work~\cite{ChatDev, AgileCoder, evomac, luo2025rpg}. 

\subsubsection{Base LLMs}
We evaluate our approach using several state-of-the-art LLMs, including three proprietary models: \textbf{\gpt{}} (gpt-4.1-20250414)~\cite{openai_gpt41_2025}, \textbf{\claudethree{}}(claude-3-5-sonnet-20241022)~\cite{anthropic_claude3.5_2024}, and \textbf{\claudefour{}} (claude-sonnet-4-20250514)~\cite{anthropic_claude_sonnet_4_2025}, as well as an open-source model, \textbf{\qwen{}} (Qwen3-Coder-480B-A35B-Instruct)~\cite{qwen3lm_qwen3_coder_2024}. 
For the proprietary models, we access them via official APIs; for the open-source model, we use the API service provided by Alibaba Cloud~\cite{aliyun_tongyi_llm_2024}.
\subsubsection{Evaluation Procedure}\label{sec:experiment:evaluation_procedure}
This section outlines the procedure for the development and evaluation of applications in \dataset{} across different \agent{}s.

\textbf{Application Development.} During the development stage, we follow the official documentation to utilize each \agent{}. 
After the user requirement and the scaffold project are configured, the agents autonomously carry out the application development. 
Due to limited resources, tasks were assigned time limits based on difficulty level: 30 minutes for elementary, 40 minutes for intermediate, and 50 minutes for advanced tasks. Development exceeding these limits will be automatically terminated. Finally, all generated applications are packaged and archived, and any logs produced by the agents (if available) are recorded for subsequent analysis.

\textbf{Application Evaluation.} A key challenge of autonomous software development is that the generated applications are unpredictable, and it remains unresolved to conduct a rigorous automatic evaluation~\cite{agent4se}. 
\revise{Therefore, following the common practice of previous studies~\cite{MetaGPT, AgileCoder}, we conduct a manual evaluation. However, as summarized in Table ~\ref{tab:benchmark}, prior studies either score applications based on overall impressions or use a binary correct/incorrect judgment for each functionality. Moreover, none of these studies accounts for evaluator bias. To overcome these limitations, we designed a more rigorous manual evaluation process, as detailed below.}

\textit{- Questionnaire Design.} \revise{To objectively measure the degree of completion for each feature}, we first design a Likert-scale questionnaire~\cite{robinson2024likert}, which comprises both functional and non-functional assessment questions.
For the \textit{functional} part, the acceptance checklist, which have been included in \dataset{}, is provided, with each requirement rated on a 4-point scale: `1' indicates the function is absent, `2' indicates the function is implemented but largely incorrect, `3' indicates the function is implemented and mostly correct, and `4' indicates the function is fully implemented and correct.
For the \textit{non-functional} assessment, we follow Creswell’s guidelines~\cite{creswell2017research} 
for designing qualitative research questions and assign the first four authors to independently identify non-functional aspects that users might encounter during evaluation. Discussions were then conducted to group and refine similar aspects until consensus was achieved. The final non-functional assessment comprises four dimensions: \textit{visual design}, \textit{usability}, \textit{stability}, and \textit{overall satisfaction}, each also rated on a 4-point scale (1 = very poor, 2 = poor, 3 = good, 4 = very good). 

\textit{- Testing Environment Construction.} To control uncertainty factors and reduce individual bias, we carefully construct a consistent testing environment for manual evaluation.
First, we assigned \textit{anonymous} identifiers to applications developed by different \agent{}s, ensuring that evaluators were unaware of which tool produced each application. After that, applications were packaged as APKs and installed on \revise{standardized} Android testing devices. Finally, we examined each application's testing prerequisites (\eg{} the Music Player and PDF Reader require local music and PDF files) and preloaded the necessary resources on the testing devices.

\textit{- Manual Evaluation.} \revise{We then recruited four evaluators from a graduate-level software testing course. After a questionnaire on acceptance-testing experience, we selected two with industrial experience and two with amateur experience. Each evaluator received a testing device with anonymized apps and a per-app questionnaire (comprising functional and non-functional assessment questions), then ran each app and completed the questionnaire.} 
After all evaluators completed the assessment, we calculated scores for each application by averaging all evaluators' scores. 

\textbf{Overall, the entire development and evaluation process takes huge
manual effort (approximately 500 person-hours) and cost (approximately 1,500 US dollars).}

\subsubsection{Metrics}
We design the following functional, non-functional, and efficiency metrics to evaluate the effectiveness and efficiency of \agent{}s in Android development.

\textbf{Functional Metrics.} 
The functional quality of developed applications is measured by: (i) \textit{Build Success Rate}: The proportion of successfully built applications. (ii) \textit{Function Completeness}: the average functional score across all applications, based on the 4-point scale defined in Section~\ref{sec:experiment:evaluation_procedure}.

\textbf{Non-functional Metrics.} The non-functional quality of the developed applications is evaluated from four perspectives: \textit{Visual Design}, \textit{Usability}, \textit{Stability}, and \textit{Overall Satisfaction}. For each dimension, we report the average score across all applications.

\textbf{Efficiency Metrics.} 
We report the average \textit{Monetary} and \textit{Time} Cost spent in development to measure the efficiency of different \agent{}s. 
In addition, inspired by prior work~\cite{MetaGPT}, we design a simple relative metric,  \textit{Productivity}, to quantify the ability of an agent to convert its cost into functional completeness, defined as:

\begin{equation}
\text{Productivity} = \frac{\text{Function Completeness} - 1}{\text{Cost}}
\end{equation}
The subtraction of 1 shifts the minimum value of Function Completeness from 1 to 0, ensuring that an agent fails to develop any runnable application yields a Productivity of 0. Higher Productivity indicates that a unit of cost produces a greater improvement in Function Completeness. We report both the \textit{Monetary Productivity} and \textit{Time Productivity} in our experimental results.
\section{Results and Analyses}
\subsection{RQ1: Effectiveness}

Table~\ref{tab:other_baselines} reports the evaluation results of \ourtool{} and other baselines with \claudefour{}.
Both open-sourced \agent{}s (\ie{} \meta{} and \gpte{}) consistently fail to generate applications that can be successfully built, which highlights the non-trivial complexity of Android development and validates the challenging nature of our constructed \dataset{} dataset. 
Claude Code achieves a high build success rate of 73.3\%. However, its Function Completeness remains low at 2.27. In practice, this means that while the generated code is syntactically valid, most functional requirements are merely superficially satisfied and largely incorrect. 
In contrast, \ourtool{} outperforms all baselines in both functional and non-functional metrics. It achieves a perfect build success rate of 100\% and the highest Function Completeness of \revise{3.57}. Additionally, all other non-functional metrics are also superior compared to the baselines.
These results demonstrate the comprehensive improvement of \ourtool{} in producing applications that are both reliable and of high quality in terms of functionality and user experience.

\finding{1}{
\ourtool{} demonstrates superior effectiveness compared to all existing LLM-agent-based approaches for software development. It consistently outperforms baselines in both functional and non-functional metrics, achieving a perfect build success rate and the highest Function Completeness score (\revise{3.57} with a \revise{\improveclaudecode{}}\% improvement compared to Claude Code).
}

\begin{table}[tb]
\centering
\caption{Evaluation results of the generated applications for different \agent{}s with \claudefour{}. }
\label{tab:other_baselines}
\begin{adjustbox}{width=0.9\textwidth}
\begin{tabular}{
    lcccccc
}
\toprule
\multirow{2}{*}{\textbf{LLM-based Agent}} &
\multicolumn{2}{c}{\textbf{Functional Metrics}} &
\multicolumn{4}{c}{\textbf{Non-Functional Metrics}} \\
\cmidrule(lr){2-3} \cmidrule(lr){4-7}
& {\makecell{\textbf{Build Success} \\ \textbf{Rate \%}}} 
& {\makecell{\textbf{Function} \\ \textbf{Completeness}}} 
& {\makecell{\textbf{Visual} \\ \textbf{Design}}} 
& {\makecell{\textbf{Usability}}} 
& {\makecell{\textbf{Stability}}} 
& {\makecell{\textbf{Overall} \\ \textbf{Satisfaction}}} \\
\midrule

\textbf{MetaGPT} &
  0.0 &
  1.00 &
  1.00 &
  1.00&
  1.00 &
  1.00 \\ 
\textbf{GPT-Engineer} &
  0.0 &
  1.00 &
  1.00 &
  1.00 &
  1.00 &
  1.00 \\ 
\midrule
\textbf{Claude Code} &
  73.3  &
  2.27 &
  2.38 &
  2.30 &
  2.25 &
  2.17 \\ \textbf{\ourtool{}} &
  \revise{\textbf{100.0 (\textcolor{red}{+26.7})}} &
  \revise{\textbf{3.57 (\textcolor{red}{+1.30})}} &
  \textbf{\revise{3.58} (\textcolor{red}{+1.20})} &
  \textbf{\revise{3.54} (\textcolor{red}{+1.24})} &
  \textbf{\revise{3.45} (\textcolor{red}{+1.20})} &
  \textbf{\revise{3.40} (\textcolor{red}{+1.23})} \\ 

\bottomrule
\end{tabular}
\end{adjustbox}
\end{table}

\subsection{RQ2: Generalizability}
To demonstrate the generalizability of \ourtool{}, we compare its performance against a single-agent setting, in which only the Programmer agent is applied to develop the entire application.

\subsubsection{RQ2.a: Model Generalizability.}
Based on the results in Table~\ref{tab:generalization}, \ourtool{} consistently improves both functional and non-functional metrics across different LLMs, with the magnitude of improvement varying by model. For the open-source \qwen{}, the single-agent setting fails to produce runnable applications, suggesting that the model might lack the fundamental coding capability required for Android development. Consequently, although integrating our framework brings improvements, its Function Completeness remains low (1.18), which indicates that most applications it develops are unusable in practice.
The Claude series demonstrates the strongest single-agent performance, with \claudethree{} reaching an average Function Completeness of 2.16 and \claudefour{} achieving an average Function Completeness of 3.07. Notably, the single-agent result of \claudefour{} even surpasses Claude Code in Table~\ref{tab:other_baselines}. 
We hypothesize that this discrepancy arises from the mismatch between Claude Code’s linear, sequential development strategy and the underlying model’s capabilities, which ultimately limit its effectiveness. \revise{For example, Claude Code often terminates early with incomplete to-do items.}
In contrast, when combined with \ourtool{}, both \claudethree{} and \claudefour{} show consistent gains across all metrics, with absolute improvements of 0.60 (27.8\% relative improvement) and \revise{0.50 (16.3}\% relative improvement) in Function Completeness.
Finally, \gpt{} demonstrates the most striking improvement when integrating with \ourtool{}, with Function Completeness rising from \revise{2.05} to 3.25 (\revise{58.5}\% relative improvement). We attribute this significant improvement to two factors. First, \gpt{} already achieves a high build success rate (100.0\%) in the single-agent setting, reflecting its ability to generate syntactically valid Android applications. However, its Function Completeness of \revise{2.05} reveals that it struggles to understand user requirements and model complex interdependencies among features. Our framework addresses this limitation by leveraging the feature map to enhance the model’s awareness of different project layers (business, design, implementation) and improve functional correctness. Second, \gpt{} benefits from its stronger instruction-following capability, which allows it to more effectively adhere to contextual guidance across iterative development. \revise{In contrast, the Claude series tends to perform extensive rounds of code review and repair, rather than leveraging the build tool to obtain direct error diagnosis. In addition, it sometimes prematurely implements functionality scheduled for later iterations, thereby disrupting the planned workflow. }

\vspace{1mm}
\begin{mdframed}[linecolor=gray,roundcorner=12pt,backgroundcolor=gray!15,linewidth=3pt,innerleftmargin=2pt, leftmargin=0cm,rightmargin=0cm,topline=false,bottomline=false,rightline = false]
    \textbf{Answer to RQ2.a:} \ourtool{} demonstrates performance gains across diverse base LLMs, consistently improving both functional and non-functional metrics, highlighting its generalization capability. Moreover, experiment results reveal that balancing the model's programming and instruction-following capabilities is crucial for iterative development, as evidenced by the greatest improvements (\revise{58.5}\% relative improvement) of \gpt{} among all base LLMs.
\end{mdframed}
\begin{table}[t]
\centering
\caption{Evaluation results of the generated applications for \ourtool{} with different LLMs.}
\label{tab:generalization}
\begin{adjustbox}{width=1.0\textwidth}

\begin{tabular}{llcccccc}

\toprule
\multirow{2}{*}{\textbf{LLM}} &
\multirow{2}{*}{\textbf{Approach}} &
\multicolumn{2}{c}{\textbf{Functional Metrics}} &
\multicolumn{4}{c}{\textbf{Non-Functional Metrics}} \\
\cmidrule(lr){3-4} \cmidrule(lr){5-8}
& & {\makecell{\textbf{Build Success} \\ \textbf{Rate \%}}} 
& {\makecell{\textbf{Function} \\ \textbf{Completeness}}} 
& {\makecell{\textbf{Visual} \\ \textbf{Design}}} 
& {\makecell{\textbf{Usability}}} 
& {\makecell{\textbf{Stability}}} 
& {\makecell{\textbf{Overall} \\ \textbf{Satisfaction}}} \\
\midrule

\multirow{2}{*}{\textbf{\qwen{}}}        & \textbf{Single Agent} & 0.0 & 1.00 & 1.00 & 1.00 & 1.00 & 1.00 \\
                                         & \textbf{\ourtool{}}  & 46.7 (\textcolor{red}{+46.7}) & 1.18 (\textcolor{red}{+0.18}) & 1.27 (\textcolor{red}{+0.27}) & 1.22 (\textcolor{red}{+0.22}) & 1.17 (\textcolor{red}{+0.17}) & 1.17 (\textcolor{red}{+0.13}) \\ \hline
\multirow{2}{*}{\textbf{\claudethree{}}} & \textbf{Single Agent} & 60.0 & 2.16 & 2.10 & 2.20 & 2.15 & 2.02 \\
                                         & \textbf{\ourtool{}}  & 93.3 (\textcolor{red}{+33.3}) & 2.76 (\textcolor{red}{+0.60}) & 2.62 (\textcolor{red}{+0.52}) & 2.57 (\textcolor{red}{+0.37}) & 2.52 (\textcolor{red}{+0.37}) & 2.42 (\textcolor{red}{+0.40}) \\ \hline
\multirow{2}{*}{\textbf{\claudefour{}}}  & \textbf{Single Agent} & 93.3 & 3.07 & 3.25 & 3.05 & 2.88 & 2.93 \\
                                         & \textbf{\ourtool{}}  & 100.0 (\textcolor{red}{+6.7}) & \revise{3.57} (\textcolor{red}{+0.50}) & \revise{3.58} (\textcolor{red}{+0.33}) & \revise{3.54} (\textcolor{red}{+0.49}) & \revise{3.45} (\textcolor{red}{+0.57}) & \revise{3.40} (\textcolor{red}{+0.47}) \\ \hline
\multirow{2}{*}{\textbf{\gpt{}}}         & \textbf{Single Agent} & \revise{100.0} & \revise{2.05} & \revise{2.03} & \revise{2.15} & \revise{2.00} & \revise{1.99} \\
                                         & \textbf{\ourtool{}}  & 100.0 (\textcolor{red}{+0.0}) & \revise{3.25} (\textcolor{red}{\textbf{+1.20}}) & \revise{3.33} (\textcolor{red}{\textbf{+1.30}}) & \revise{3.22} (\textcolor{red}{\textbf{+1.07}}) & \revise{3.12} (\textcolor{red}{\textbf{+1.12}}) & \revise{3.02} (\textcolor{red}{\textbf{+1.03}}) \\

\bottomrule
\end{tabular}

\end{adjustbox}
\end{table}

\subsubsection{RQ2.b: Task Generalizability.}
As shown in Table~\ref{tab:difficulty}, \ourtool{} consistently outperforms the single agent across tasks of varying difficulty, demonstrating strong generalizability.
For Elementary tasks, the baseline performance is already high; however, after optimization, the Function Completeness score reaches near-optimal levels (\eg{} \revise{3.79}/4 for \claudefour{}). 
The Intermediate tasks in \dataset{} appear to lie near the capability boundary of single agents. As an illustrative example, \claudethree{} fails on six tasks in the single-agent setting, five of which rank among the most challenging in \dataset{}, suggesting that these tasks require reasoning and decomposition capabilities that surpass what a single agent can provide.
At this difficulty level, \ourtool{} effectively leverages the feature map to guide task decomposition and manage feature dependencies, enabling the agent to get necessary contexts from the business, design, and implementation layers, thereby achieving remarkable gains.
\revise{For Advanced tasks, while the Claude series exhibits relatively small improvements in Function Completeness, \gpt{} still achieves a substantial gain, improving from 1.30 to 2.64.
Further case analysis reveals that this difference might be partly explained by task-specific effects. In particular, the \textit{Shoot'em Up} application, a classic video game, appears to be overfitted by the Claude series since their single-agent baselines achieve unusually high scores on this task (\eg{} \claudethree{} reaches 3.76 compared to an average of 1 on other Advanced apps, and \claudefour{} achieves 3.30 compared to 2.82). This overfitting effect elevates their single-agent performance on Advanced tasks.} However, \ourtool{} still delivers consistent improvements by providing structured decomposition and dependency-aware context assistance.

\vspace{1mm}
\begin{mdframed}[linecolor=gray,roundcorner=12pt,backgroundcolor=gray!15,linewidth=3pt,innerleftmargin=2pt, leftmargin=0cm,rightmargin=0cm,topline=false,bottomline=false,rightline = false]
    \textbf{Answer to RQ2.b:} \ourtool{} demonstrates robust and consistent improvements across all difficulty levels, especially for more complex development tasks, where single agents often reach their capability boundary and fail to capture complex dependencies among user requirements. These results highlight the adaptability and effectiveness of \ourtool{} across diverse task settings.
\end{mdframed}
\begin{table}[t]
\centering
\caption{Evaluation results of the generated applications for \ourtool{} across tasks of varying difficulty.}
\label{tab:difficulty}
\begin{adjustbox}{width=1.0\textwidth}

\begin{tabular}{lllcccccc}

\toprule
\multirow{2}{*}{\textbf{LLM}} &
\multirow{2}{*}{\textbf{Approach}} &
\multirow{2}{*}{\textbf{Difficulty}} &
\multicolumn{2}{c}{\textbf{Functional Metrics}} &
\multicolumn{4}{c}{\textbf{Non-Functional Metrics}} \\
\cmidrule(lr){4-5} \cmidrule(lr){6-9}
& & & {\makecell{\textbf{Build Success} \\ \textbf{Rate \%}}} 
& {\makecell{\textbf{Function} \\ \textbf{Completeness}}} 
& {\makecell{\textbf{Visual} \\ \textbf{Design}}} 
& {\makecell{\textbf{Usability}}} 
& {\makecell{\textbf{Stability}}} 
& {\makecell{\textbf{Overall} \\ \textbf{Satisfaction}}} \\
\midrule

\multirow{8}{*}{\textbf{\claudethree{}}} &
  \multirow{4}{*}{\textbf{Single Agent}} &
  Elementary &
  85.7 &
  2.78 &
  2.57 &
  2.71 &
  2.86 &
  2.57 \\
 &
   &
  Intermediate &
  40.0 &
  1.42 &
  1.75 &
  1.70 &
  1.45 &
  1.40 \\
 &
   &
  Advanced &
  33.3 &
  1.92 &
  1.58 &
  1.83 &
  1.67 &
  1.75 \\
 &
   &
  All &
 \cellcolor[HTML]{EFEFEF} 60.0 &
  \cellcolor[HTML]{EFEFEF}2.16 &
  \cellcolor[HTML]{EFEFEF}2.10 &
  \cellcolor[HTML]{EFEFEF}2.20 &
  \cellcolor[HTML]{EFEFEF}2.15 &
  \cellcolor[HTML]{EFEFEF}2.02 \\ \cline{2-9} 
 &
  \multirow{4}{*}{\textbf{\ourtool{}}} &
  Elementary &
  100.0 (\textcolor {red}{+14.3}) &
  3.22 (\textcolor {red}{+0.44}) &
  2.93 (\textcolor {red}{+0.36}) &
  3.00 (\textcolor {red}{+0.29}) & 
  2.96 (\textcolor {red}{+0.10}) &
  2.79 (\textcolor {red}{+0.22}) \\
  & & Intermediate &
  80.0 (\textcolor {red}{+40.0}) &
  2.50 (\textcolor {red}{+1.08}) &
  2.45 (\textcolor {red}{+0.70}) &
  2.25 (\textcolor {red}{+0.55}) & 
  2.25 (\textcolor {red}{+0.80}) & 
  2.15 (\textcolor {red}{+0.75}) \\
  & & Advanced &
  100.0 (\textcolor {red}{+66.7}) &
  2.10 (\textcolor {red}{+0.18}) &
  2.17 (\textcolor {red}{+0.59}) &
  2.08 (\textcolor {red}{+0.25}) & 
  1.92 (\textcolor {red}{+0.25}) & 
  2.00 (\textcolor {red}{+0.25}) \\
  & & All &
  \cellcolor [HTML]{EFEFEF} 93.3 (\textcolor {red}{+33.3}) & 
  \cellcolor [HTML]{EFEFEF} 2.76 (\textcolor {red}{+0.60}) & 
  \cellcolor [HTML]{EFEFEF} 2.62 (\textcolor {red}{+0.52}) & 
  \cellcolor [HTML]{EFEFEF} 2.57 (\textcolor {red}{+0.37}) & 
  \cellcolor [HTML]{EFEFEF} 2.52 (\textcolor {red}{+0.37}) & 
  \cellcolor [HTML]{EFEFEF} 2.42 (\textcolor {red}{+0.40}) \\ \hline
  \multirow{8}{*}{\textbf{\claudefour{}}} &
  \multirow{4}{*}{\textbf{Single Agent}} &
  Elementary &
  100.0 &
  3.35 &
  3.57 &
  3.28 &
  3.18 &
  3.21 \\
 &
   &
  Intermediate &
  80.0 &
  2.75 &
  2.85 &
  2.80 &
  2.60 &
  2.70 \\
 &
   &
  Advanced &
  100.0 &
  2.96 &
  3.17 &
  2.92 &
  2.67 &
  2.67 \\
 &
   &
  All &
  \cellcolor[HTML]{EFEFEF}93.3 &
  \cellcolor[HTML]{EFEFEF}3.07 &
  \cellcolor[HTML]{EFEFEF}3.25 &
  \cellcolor[HTML]{EFEFEF}3.05 &
  \cellcolor[HTML]{EFEFEF}2.88 &
  \cellcolor[HTML]{EFEFEF}2.93 \\ \cline{2-9} 
 &
  \multirow{4}{*}{\textbf{\ourtool{}}} &
  Elementary &
100.0 (\textcolor {red}{+0.0}) & 
\revise{3.79} (\textcolor {red}{+0.44}) & 
\revise{3.71} (\textcolor {red}{+0.14}) & 
\revise{3.71}(\textcolor {red}{+0.43}) & 
\revise{3.71} (\textcolor {red}{+0.53}) & 
\revise{3.54} (\textcolor {red}{+0.33}) \\ 
& & Intermediate &
100.0 (\textcolor {red}{+20.0}) & 
\revise{3.62} (\textcolor {red}{+0.87}) & 
\revise{3.58} (\textcolor {red}{+0.73}) & 
\revise{3.58} (\textcolor {red}{+0.78}) & 
\revise{3.42} (\textcolor {red}{+0.82}) & 
\revise{3.46} (\textcolor {red}{+0.76}) \\ 
& & Advanced &
100.0 (\textcolor {red}{+0.0}) & 
3.03 (\textcolor {red}{+0.07}) & 
3.33 (\textcolor {red}{+0.16}) & 
3.08 (\textcolor {red}{+0.16}) & 
3.00 (\textcolor {red}{+0.33}) & 
3.00 (\textcolor {red}{+0.33}) \\ 
& & All &
\cellcolor [HTML]{EFEFEF} 100.0 (\textcolor {red}{+6.7}) & 
\cellcolor [HTML]{EFEFEF} \revise{3.57} (\textcolor {red}{+0.50}) & 
\cellcolor [HTML]{EFEFEF} \revise{3.58} (\textcolor {red}{+0.33}) & 
\cellcolor [HTML]{EFEFEF} \revise{3.54} (\textcolor {red}{+0.49}) & 
\cellcolor [HTML]{EFEFEF} \revise{3.45} (\textcolor {red}{+0.57}) & 
\cellcolor [HTML]{EFEFEF} \revise{3.40} (\textcolor {red}{+0.47}) \\ \hline

\multirow{8}{*}{\textbf{\gpt{}}} &
  \multirow{4}{*}{\textbf{Single Agent}} &
  Elementary &
  100.0 &
  \revise{3.09} &
  \revise{2.92} &
  \revise{3.12} &
  \revise{2.96} &
  \revise{2.79} \\
 &
   &
  Intermediate &
  \revise{100.0} &
  \revise{1.39} &
  \revise{1.25} &
  \revise{1.50} &
  \revise{1.25} &
  \revise{1.33} \\
 &
   &
  Advanced &
  100.0 &
  \revise{1.30} &
  \revise{1.54} &
  \revise{1.50} &
  \revise{1.42} &
  \revise{1.50} \\
 &
   &
  All &
  \cellcolor[HTML]{EFEFEF} \revise{100.0} &
\cellcolor[HTML]{EFEFEF} \revise{2.05} &
  \cellcolor[HTML]{EFEFEF} \revise{2.03} &
  \cellcolor[HTML]{EFEFEF} \revise{2.15} &
  \cellcolor[HTML]{EFEFEF} \revise{2.00} &
  \cellcolor[HTML]{EFEFEF} \revise{1.99} \\ \cline{2-9} 
 &
  \multirow{4}{*}{\textbf{\ourtool{}}} &
  Elementary &
100.0 (\textcolor {red}{+0.0}) & 
\revise{3.62} (\textcolor {red}{+0.53}) & 
\revise{3.46} (\textcolor {red}{+0.54}) & 
\revise{3.50} (\textcolor {red}{+0.38}) & 
\revise{3.58} (\textcolor {red}{+0.62}) & 
\revise{3.42} (\textcolor {red}{+0.63}) \\
& & Intermediate &
100.0 (\textcolor {red}{+0.0}) & 
\revise{3.20} (\textcolor {red}{+1.81}) & 
\revise{3.29} (\textcolor {red}{+2.04}) &
\revise{3.17} (\textcolor {red}{+1.67}) & 
\revise{2.92} (\textcolor {red}{+1.67}) & 
\revise{2.79} (\textcolor {red}{+1.46}) \\
& & Advanced &
100.0 (\textcolor {red}{+0.0}) & 
\revise{2.64} (\textcolor {red}{+1.34}) & 
\revise{3.17} (\textcolor {red}{+1.63}) & 
\revise{2.75} (\textcolor {red}{+1.25}) & 
\revise{2.58} (\textcolor {red}{+1.16}) &
\revise{2.67} (\textcolor {red}{+1.17}) \\
& & All &
\cellcolor [HTML]{EFEFEF} 100.0 (\textcolor {red}{+0.0}) & 
\cellcolor [HTML]{EFEFEF} \revise{3.25} (\textcolor {red}{+1.20}) & 
\cellcolor [HTML]{EFEFEF} \revise{3.33} (\textcolor {red}{+1.30}) & 
\cellcolor [HTML]{EFEFEF} \revise{3.22} (\textcolor {red}{+1.07}) & 
\cellcolor [HTML]{EFEFEF} \revise{3.12} (\textcolor {red}{+1.12}) & 
\cellcolor [HTML]{EFEFEF} \revise{3.02} (\textcolor {red}{+1.03}) \\

\bottomrule
\end{tabular}

\end{adjustbox}
\end{table}
\subsection{RQ3: Ablation Study}
\revise{Table~\ref{tab:ablation} demonstrates the contributions of different strategies in \ourtool{} to the final performance improvements. 
Despite \claudefour{} already achieving the strongest single-agent performance (\ie{} 3.07 Function Completeness), incorporating these strategies consistently improves outcomes. 
Specifically, introducing the overall design, which establishes a coherent blueprint of the app’s structure and guides subsequent development, results in an absolute improvement of 0.22 (7.2\% relative improvement) in Function Completeness.
}
\revise{Building on this, we introduce the feature map to conduct feature decomposition. We first experimented with removing the context of predecessor features across iterations, leaving the agent to develop merely based on the current feature description. Results indicate that simply replacing single-pass development with iterative development, without effective context management, leads to a notable drop in both Build Success rate (from 100.0\% to 86.7\%) and Function Completeness (from 3.29 to 2.80). This is because iterative development requires the agent to repeatedly retrieve relevant context and locate extension points across the entire repository, which has been widely recognized as a highly challenging task~\cite{lingma, rahardja2025can}.
}
\revise{
In contrast, when access to the contexts of predecessor features, the programmer agents can make more accurate coding decisions, which confers a genuine advantage to iterative development (an absolute improvement of \revise{0.50}, a 16.3\% relative gain in Function Completeness).
\revise{For example, in the \textit{Expense Tracker} application, the expenditure needs to be aggregated by date. Without predecessor contexts, the agent writes down \texttt{expenses.sumOf\{it.amount\}}. However, \texttt{expenses} is actually a collection of \texttt{ExpenseWithCategory} entities, and the correct way to access the expenditure is \texttt{it.expense.amount}.  This mistake finally leads to development failures. In contrast, the modifications to \texttt{ExpenseWithCategory} are explicitly propagated to the next iteration in \ourtool{}, enabling the agent to correctly recognize the nested data structure and produce the correct implementation.}
}

\vspace{1mm}
\begin{mdframed}[linecolor=gray,roundcorner=12pt,backgroundcolor=gray!15,linewidth=3pt,innerleftmargin=2pt, leftmargin=0cm,rightmargin=0cm,topline=false,bottomline=false,rightline = false]
    \textbf{Answer to RQ3:} 
    \revise{
    \ourtool{} provides a practical method for mitigating the inherent difficulties of retrieving relevant context and locating extension points in iterative development, with both the overall design and feature map-guided iterative development contributing  to the overall performance, yielding gains of 0.22 (7.2\% relative improvement) and \revise{0.50} (\revise{16.3}\% relative improvement) in Function Completeness, on the best-performing \claudefour{} model.} 
\end{mdframed}

\begin{table}[t]
\centering
\caption{Ablation study of \ourtool{} with \claudefour{}.}
\label{tab:ablation}
\begin{adjustbox}{width=1.0\textwidth}
\begin{tabular}{cccccccccc}

\toprule
\multirow{3}{*}{\makecell{\textbf{Overall Design} \\ \textbf{Construction}}} &
\multirow{3}{*}{\makecell{\textbf{Feature Map} \\ \textbf{Generation}}} &
\multirow{3}{*}{\makecell{\revise{\textbf{Predecessor}} \\ \revise{\textbf{Contexts}}}} &
\multirow{3}{*}{\makecell{\textbf{Iterative/Single-pass} \\ \textbf{Development}}}  &
\multicolumn{2}{c}{\textbf{Functional Metrics}} &
\multicolumn{4}{c}{\textbf{Non-Functional Metrics}} \\
\cmidrule(lr){5-6} \cmidrule(lr){7-10}
& & & & {\makecell{\textbf{ Build Success } \\ \textbf{Rate \%}}} 
& {\makecell{\textbf{Function} \\ \textbf{Completeness}}} 
& {\makecell{\textbf{Visual} \\ \textbf{Design}}} 
& {\makecell{\textbf{Usability}}} 
& {\makecell{\textbf{Stability}}} 
& {\makecell{\textbf{Overall} \\ \textbf{Satisfaction}}} \\
\midrule
\redCross & \redCross  & \redCross & \revise{Single-pass}   & 93.3 & 3.07 & 3.25 & 3.05 & 2.88 & 2.93 \\

\greenCheck & \redCross & \redCross & \revise{Single-pass}  & 100.0 (\textcolor{red}{+6.7}) &  3.29 (\textcolor{red}{+0.22}) & 3.43 (\textcolor{red}{+0.18}) & 3.33 (\textcolor{red}{+0.28}) & 3.25 (\textcolor{red}{+0.37}) & 3.22 (\textcolor{red}{+0.29}) \\

\greenCheck & \greenCheck  & \redCross & \revise{Iterative}  & \revise{86.7} (\textcolor{upgreen}{-6.6}) & \revise{2.80} (\textcolor{upgreen}{-0.27}) & \revise{3.01} (\textcolor{upgreen}{-0.24}) & \revise{2.72} (\textcolor{upgreen}{-0.33}) & \revise{2.65} (\textcolor{upgreen}{-0.23}) & \revise{2.52} (\textcolor{upgreen}{-0.41}) \\

\greenCheck & \greenCheck & \greenCheck & \revise{Iterative}  & 100.0 (\textcolor{red}{+6.7}) & \revise{3.57} (\textcolor{red}{+0.50}) & \revise{3.58} (\textcolor{red}{+0.33}) & \revise{3.54} (\textcolor{red}{+0.49}) & \revise{3.45} (\textcolor{red}{+0.57}) & \revise{3.40} (\textcolor{red}{+0.47}) \\

\bottomrule
\end{tabular}
\end{adjustbox}
\end{table}

\subsection{RQ4: Effiency}
Table~\ref{tab:efficiency} presents the monetary and time cost for different \agent{}s. First, on both \gpt{} and \claudethree{}, our approach achieves comparable monetary and time costs to the single-agent baselines. In addition, the higher Productivity metrics indicate that although our tool introduces extra overhead, it yields higher function completeness under the same cost. For instance, on \gpt{}, we incur an extra \revise{0.38} dollars and around \revise{2} minutes per application, yet achieve a \revise{\improvesingle{}}\% improvement in Function Completeness, demonstrating the effectiveness of our method.
We attribute this improvement to two factors. First, the single agent retains the entire development context in its dialogue history, which grows with project size and increases token costs. In contrast, our approach stores hierarchical information in the feature map and provides only the relevant context for each iteration, reducing overhead caused by cluttered contexts. Feature decomposition also simplifies the difficulty of implementation, decreasing the time spent on repeated debugging. Second, as described in Section~\ref{sec:method:feature_developement}, we optimize the single-agent trajectory to prevent dialogue histories from being polluted with fragmented code modifications and redundant file versions, further reducing token costs. 

The only exception occurs in experiments with \claudefour{} as the base LLM, \revise{where the productivity of all LLM-based agents is significantly lower than that of the single-agent setting.} We assume that this deviation stems from \claudefour{}’s intrinsic development behavior.
As we mentioned in RQ2.a, \claudefour{} tends to perform multiple rounds of code review and repair in development, which improves the single-pass generation accuracy but increases time and token usage. 
To validate this hypothesis, we randomly select two applications (\ie{} \textit{Shopping List} and \textit{Countdown Timer}) and manually check the number of development rounds of both \gpt{} and \claudefour{}. The result shows that \claudefour{} performs 12 rounds per feature, while GPT-4.1 only performs 5 rounds.

We further analyze the open-source baselines, \meta{} and \gpte{}, which both fail to produce any runnable applications. Among all the approaches, \meta{} incurs the highest overhead (\$9.61 and around 24 minutes per app). This might be due to its excessive design; for example, even for the elementary \textit{QRCode Tool} app, it generated 47 files and over 7,000 lines of code. In contrast, GPT-Engineer performs minimal design and code modification, with an average number of code lines being less than 350. This demonstrates that even with state-of-the-art LLMs, the design of the development workflow critically affects outcomes, with overly complex or overly simple designs both leading to failure. Our approach strikes a balance between workflow design and code implementation, consistently improving performance across different scenarios.

\vspace{1mm}
\begin{mdframed}[linecolor=gray,roundcorner=12pt,backgroundcolor=gray!15,linewidth=3pt,innerleftmargin=2pt, leftmargin=0cm,rightmargin=0cm,topline=false,bottomline=false,rightline = false]
   \textbf{Answer to RQ4:} On \gpt{} and \claudethree{},  \ourtool{} achieves comparable efficiency performance to single-agent baselines while delivering higher Function Completeness per unit cost. However, the intrinsic agent behaviors of \claudefour{} make it tend to perform fine-grained review and fix operations, which require more modification rounds per feature. Iterative development further amplifies this issue and incurs more cost, highlighting the need to balance the model-intrinsic behaviors with the guidance of the external workflow.
\end{mdframed}
\begin{table}[t]
\centering
\caption{Average cost of different \agent{}s in developing applications in \dataset{}}
\label{tab:efficiency}
\begin{adjustbox}{width=0.8\textwidth}
\begin{tabular}{llrrrrr}

\toprule

\multirow{2}{*}{\textbf{LLM}} &
  \multirow{2}{*}{\textbf{Approach}} &
  \multicolumn{2}{c}{\textbf{Monetary Metrics}} & &
  \multicolumn{2}{c}{\textbf{Time Metrics}} \\ \cline{3-4} \cline{6-7} 
 &
   &
  \textbf{Cost (\$)} &
  \multicolumn{1}{r}{\textbf{Productivity}} & &
  \textbf{Cost (min)} &
  \multicolumn{1}{r}{\textbf{Productivity}} \\ 
  
\midrule
\multirow{2}{*}{\textbf{\gpt{}}}        & \textbf{Single Agent} & \revise{0.88}                & \revise{1.20} & & \revise{8.23} & \revise{0.13} \\
                                        & \textbf{\ourtool{}}        & \revise{1.26}                & \revise{1.79} \textcolor{red}{($\uparrow$ 49.2\%)} & & \revise{10.37}  & \revise{0.22} \textcolor{red}{($\uparrow$ 69.2\%)} \\ \hline
\multirow{2}{*}{\textbf{\claudethree{}}} &
  \textbf{Single Agent} &
  2.07 &
  0.58 & &
  14.52 &
  0.08 \\
                                        & \textbf{\ourtool{}}        & 2.88                & 0.61 \textcolor{red}{($\uparrow$ 5.2\%)} & & 18.23 & 0.10 \textcolor{red}{($\uparrow$ 25.0\%)} \\ \hline

\multirow{5}{*}{\textbf{\claudefour{}}}
                                        & \textbf{Single Agent} & 1.56                & 1.37 & & 9.18  & 0.23 \\
                                        & \textbf{MetaGPT}           & 9.61                & 0.00 \textcolor{upgreen}{($\downarrow$ 100.0\%)} & & 23.84 & 0.00 \textcolor{upgreen}{($\downarrow$ 100.0\%)} \\
                                        & \textbf{GPT-Engineer}      & 0.09                & 0.00 \textcolor{upgreen}{($\downarrow$ 100.0\%)} & & 1.28  & 0.00 \textcolor{upgreen}{($\downarrow$ 100.0\%)}\\
                                        
                                        & \textbf{Claude Code}       & 4.38                & 0.27 \textcolor{upgreen}{($\downarrow$ 80.3\%)} & & 11.81 & 0.10 \textcolor{upgreen}{($\downarrow$ 56.5\%)} \\
                                        & \textbf{\ourtool{}}        & \revise{4.12}                & \revise{0.62} \textcolor{upgreen}{($\downarrow$ 54.7\%)} & & \revise{23.19} & \revise{0.11} \textcolor{upgreen}{($\downarrow$ 52.2\%)} \\ 
\bottomrule

\end{tabular}

\end{adjustbox}
\end{table}
\revise{\subsection{RQ5: Failure Analysis}}
\begin{figure}[htbp]
\centering
\begin{minipage}[t]{0.55\textwidth}
\centering
\captionof{table}{\revise{Taxonomy of common failure modes of LLM-based agents in end-to-end software development.}}
\label{tab:failure_tax}
\begin{adjustbox}{width=1.0\textwidth}
\begin{tabular}{>{\centering\arraybackslash}p{2.7cm}lp{5cm}}
\toprule
\textbf{Failure Category} & \textbf{Sub-Category} & \textbf{Description} \\ \midrule
\multirow{5}{*}{\makecell[c]{\textbf{Procedural} \\ \textbf{Failures}}}
 &  Requirement Omission & Omitting user requirements. \\ \cmidrule{2-3}

 & \multirow{2}{*}{Incomplete Procedure} & Early stop or Missing iterative build and fix steps. \\ \cmidrule{2-3}

 & \multirow{2}{*}{Wrong Software Type} & Developing web applications instead of Android. \\
 \cmidrule{2-3}
 & \multirow{1}{*}{Execution Timeout} & Timeout due to a fix loop. \\
 \midrule
\multirow{3}{*}{\makecell[c]{\textbf{Contextual} \\ \textbf{Failures}}} & \multirow{3}{*}{Interface Omission} & Omitting inter-layer interface, for example, implementing logic methods without corresponding UI. \\ 

\midrule

\multirow{9}{*}{\makecell[c]{\textbf{Implementation} \\ \textbf{Failures}}} & \multirow{2}{*}{Missing  Configuration} & Missing required libraries or permission requests. \\ \cmidrule{2-3}
 & \multirow{3}{*}{Wrong API Usage} & Specific APIs were used incorrectly, for example, Hardware bitmaps are not accessible from the CPU. \\ \cmidrule{2-3}
& \multirow{2}{*}{Faulty Logic} & The function logic is faulty, for example, the UI does not update. \\
 \bottomrule

\end{tabular}

\end{adjustbox}
\end{minipage}
\hfill
\begin{minipage}[t]{0.42\textwidth}
\vspace{3pt}
\centering
\includegraphics[width=0.95\linewidth]{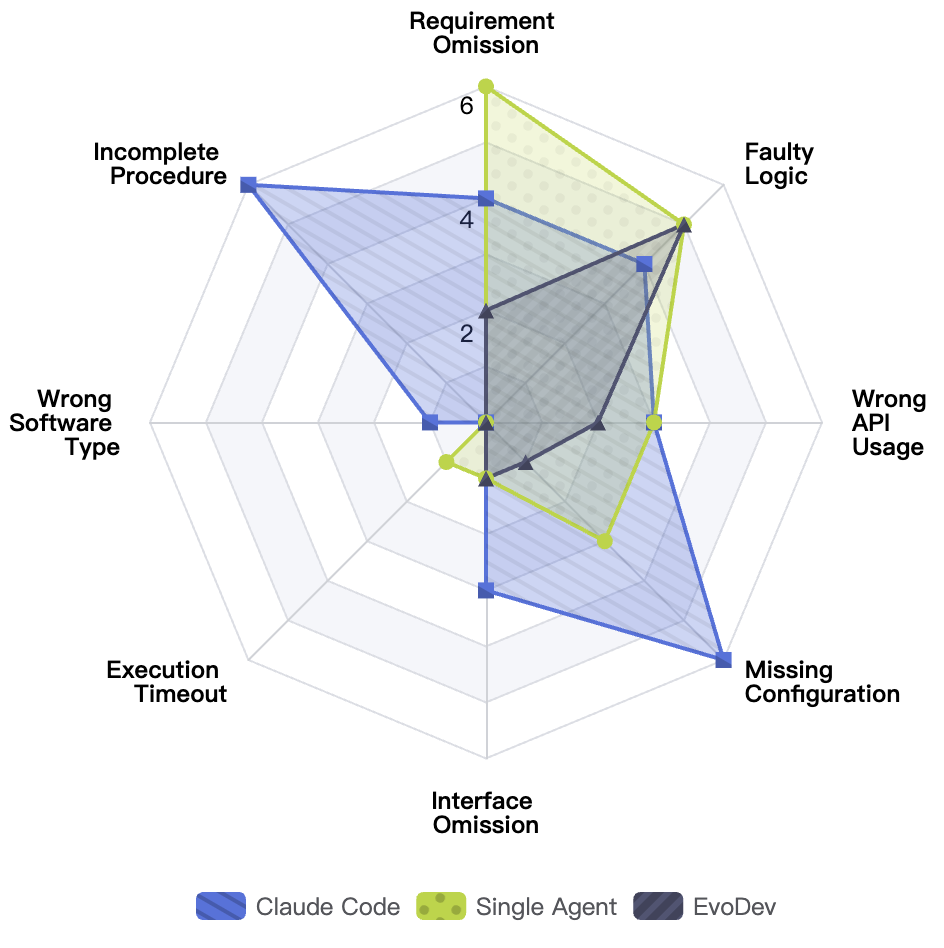}
\caption{Distribution of failure modes of Claude Code, Single Agent, and EvoDev.}
\label{fig:failure}
\end{minipage}
\end{figure}

\revise{
Table ~\ref{tab:failure_tax} summarizes the failure modes observed in end-to-end software development across different agents. \textit{Procedural Failures} arise from incorrect or incomplete development procedures. \textit{Contextual Failures} occur when an agent ignores necessary code context from previous iterations (\eg{} inter-layer interfaces). \textit{Implementation Failures }result from mistakes during feature implementation.
}
\revise{
Figure ~\ref{fig:failure} compares the distribution of failure modes across Claude Code, Single Agent, and EvoDev.
}
\revise{
Among all agents, Claude Code exhibits the largest number and variety of failures in end-to-end software development. Although Claude Code is more versatile, it grants the LLM excessive freedom, resulting in a less controlled development process. Its to-do lists are often incomplete, missing requirements or necessary steps (\eg{} invoking the build tool); it often stops prematurely with uncompleted to-do items; it retains only a one-sentence summary per iteration, causing the agent to merely focus on core logic while neglecting necessary configurations such as third-party libraries or permission requests.
}
\revise{
Single Agent mainly suffers from requirement omission and implementation failures arising from developing the entire application in a single pass. This illustrates the limitations of single-pass development: as application scale increases, the agent is prone to requirement omissions and implementation failures. Single Agent also experienced a case of timeout due to a fix loop, showing performance degradation under long contexts.
}
\revise{
In contrast, \ourtool{} enhances Single Agent with global design, feature-map guided task decomposition, and context propagation, substantially mitigating its issues. Both the number and severity of failures are significantly lower than those of other baselines. The only exception is \textit{Faulty Logic}, where no clear decrease is observed. This is expected since our method mainly improves high-level feature decomposition and context propagation, while the implementation of atomic features still depends on the LLM’s inherent coding ability. Failures may still occur in complex features, such as ``drag the magnifier'' in the \textit{Color Picker} application.
}
\vspace{1mm}
\begin{mdframed}[linecolor=gray,roundcorner=12pt,backgroundcolor=gray!15,linewidth=3pt,innerleftmargin=2pt, leftmargin=0cm,rightmargin=0cm,topline=false,bottomline=false,rightline = false]
   \textbf{Answer to RQ5:} 
   \ourtool{} substantially reduces both the frequency and severity of different failures through global design, feature-map-guided planning, and context propagation. Nevertheless, failures may still occur for complex features that rely primarily on the underlying LLM’s capabilities, which is a common challenge for all software development agents.
\end{mdframed}


\section{Dicussion}
\subsection{Implications}
We summarize the following insightful implications based on the findings in our experiments.

\textbf{Iterative Development Rewards Balanced Capabilities.} 
The coding capability is regarded as the key metric to measure the performance of LLMs in software engineering tasks. However, as we revealed in RQ2.a, while the Claude series demonstrates superior single-agent coding capabilities, \gpt{} benefits far more from the iterative development framework due to its stronger global planning and instruction-following capabilities.
Experimental results on \qwen{} also suggest that the coding capability remains a bottleneck. This highlights a tradeoff between enhancing the coding capabilities and instruction-following capabilities of LLMs.

\textbf{Iterative Development Demands Context Management.} As explored in RQ3, merely splitting features across iterations without providing the necessary context to support their evolution can significantly degrade development performance, even performing worse than single-agent single-pass development. This highlights that effective context management is crucial in iterative development, especially the contextual dependencies between iterations.

\textbf{Model-Intrinsic Behaviors May Misalign Workflow Guidance.}
Our experiments across different LLM-agent-based baselines reveal a potential mismatch between model-intrinsic and workflow-guided behaviors.  
As described in RQ4, \claudefour{} insisted on a multi-round self-editing and revision behavior pattern, even when we set explicit prompting constraints (\eg{} avoiding self-revision and merely fixing errors based on the build feedback). This behavior pattern strengthens its coding success rate, but sometimes also disturbs the workflow plan and incurs more overhead.
We hypothesize that this might be attributed to its overfitting to such a behavior pattern during its training to achieve higher coding performance. However, such model-intrinsic agent behaviors can conflict with those expected by the external workflows, demonstrating the misalignment of training more autonomous agents with the design of downstream agent systems.

\textbf{Iterative Development Can Be Cost-Efficient.}
Iterative workflows intuitively appear more costly and less efficient. However, our experiments show that when applied to \gpt{} and \claudethree{}, the additional overhead remains within acceptable bounds. Moreover, when measuring the function completeness gained by unit cost, iterative approaches can even surpass single-pass generation in efficiency. Therefore, iterative development can also be cost-efficient and might be a promising future paradigm for sustainable AI-assisted software engineering.

\subsection{Threats to Validity}

\textit{Internal Threats.}
The first internal threat lies in our manual evaluation of the developed applications, which might incur individual biases and mistakes. To mitigate this, we carefully design a Likert-scale questionnaire to assist in collecting objective feedback and invite four participants to calculate the average scores.
\revise{The second threat arises from variations in user input. Minor perturbations in natural language requirements can lead to a vast and unstructured exploration space, making exhaustive manual evaluation infeasible. As a result, a comprehensive analysis of input sensitivity remains impractical in the absence of fully automated end-to-end evaluation methods. We introduce a Business Analyst agent to partially mitigate this issue, but fine-grained requirement elicitation and clarification remain a challenge for future research.}

\textit{External Threats.}
There are some threats that impact the generalizability of our study. First, our dataset is of a limited size due to the extensive evaluation cost. However, we try our best to design a rigorous dataset construction process. These efforts mitigate the risks of bias and improve the objectivity and applicability of our dataset.
Second, our implementation and evaluation of \ourtool{} are specific to Android development with Kotlin. As a result, the findings may not be generalizable to other types of applications and programming languages. Exploring the iterative software development process for different languages is a valuable direction for future research.

\subsection{Limitations}
We then discuss the limitations of our framework. First, we do not incorporate human-in-the-loop interaction in the current approach. Nevertheless, our framework provides multiple extension points for such interaction. For example, users can inspect and modify the feature map, and after each iteration, users can input new requirements that can be integrated into the feature map. We plan to incorporate this capability in future work. 
Second, our method does not include testing. Our preliminary experiments indicate that integrating testing into Android development remains challenging for LLM-based agents. These agents struggle to generate correct test code, and when test failures occur, they have difficulty determining whether the faults stem from the test code or the code under test, which increases the likelihood of failures. Integrating an effective testing mechanism in iterative development is a promising research direction.

\section{Conclusion}
This work proposes \ourtool{}, a novel iterative framework for end-to-end software development. Inspired by classic feature-driven development methodology, \ourtool{} constructs a feature map to store the context of each feature and manages complex dependencies among them, which helps provide necessary contexts for each iteration. Experimental results show that \ourtool{} not only surpasses existing baselines but also significantly improves the performance of single agents across different base LLMs, with an acceptable overhead and higher productivity. Further analysis reveals insightful implications for the demands that iterative development places on LLM capabilities and context management, and raises the
concern about the misalignment of agent-intrinsic with framework-guided behaviors.



\bibliographystyle{ACM-Reference-Format}
\bibliography{ref}

@inproceedings{MetaGPT,
  author       = {Sirui Hong and
                  Mingchen Zhuge and
                  Jonathan Chen and
                  Xiawu Zheng and
                  Yuheng Cheng and
                  Jinlin Wang and
                  Ceyao Zhang and
                  Zili Wang and
                  Steven Ka Shing Yau and
                  Zijuan Lin and
                  Liyang Zhou and
                  Chenyu Ran and
                  Lingfeng Xiao and
                  Chenglin Wu and
                  J{\"{u}}rgen Schmidhuber},
  title        = {MetaGPT: Meta Programming for {A} Multi-Agent Collaborative Framework},
  booktitle    = {The Twelfth International Conference on Learning Representations,
                  {ICLR} 2024, Vienna, Austria, May 7-11, 2024},
  publisher    = {OpenReview.net},
  year         = {2024},
  url          = {https://openreview.net/forum?id=VtmBAGCN7o},
  bibsource    = {dblp computer science bibliography, https://dblp.org}
}

@inproceedings{ChatDev,
  author       = {Chen Qian and
                  Wei Liu and
                  Hongzhang Liu and
                  Nuo Chen and
                  Yufan Dang and
                  Jiahao Li and
                  Cheng Yang and
                  Weize Chen and
                  Yusheng Su and
                  Xin Cong and
                  Juyuan Xu and
                  Dahai Li and
                  Zhiyuan Liu and
                  Maosong Sun},
  editor       = {Lun{-}Wei Ku and
                  Andre Martins and
                  Vivek Srikumar},
  title        = {ChatDev: Communicative Agents for Software Development},
  booktitle    = {Proceedings of the 62nd Annual Meeting of the Association for Computational
                  Linguistics (Volume 1: Long Papers), {ACL} 2024, Bangkok, Thailand,
                  August 11-16, 2024},
  pages        = {15174--15186},
  publisher    = {Association for Computational Linguistics},
  year         = {2024},
  url          = {https://doi.org/10.18653/v1/2024.acl-long.810},
  doi          = {10.18653/V1/2024.ACL-LONG.810},
  bibsource    = {dblp computer science bibliography, https://dblp.org}
}

@Manual{lovable,
  author = {Lovable},
  title =        {Lovable},
  note =         {\url{https://lovable.dev/}},
  year=2025
}

@Manual{GPT-Engineer,
  author = {AntonOsika},
  title =        {GPT-Engineer},
  note =         {\url{https://github.com/AntonOsika/gpt-engineer}},
  year=2025
}

@Manual{Claude-Code,
  author = {Anthropic},
  title =        {Claude Code},
  note =         {\url{https://www.anthropic.com/claude-code}},
  year=2025
}

@book{creswell2017research,
  title={Research design: Qualitative, quantitative, and mixed methods approaches},
  author={Creswell, John W and Creswell, J David},
  year={2017},
  publisher={Sage publications}
}

@inproceedings{AgileCoder,
  author       = {Minh Huynh Nguyen and
                  Thang Phan Chau and
                  Phong X. Nguyen and
                  Nghi D. Q. Bui},
  title        = {AgileCoder: Dynamic Collaborative Agents for Software Development
                  based on Agile Methodology},
  booktitle    = {{IEEE/ACM} Second International Conference on {AI} Foundation Models
                  and Software Engineering, Forge@ICSE 2025, Ottawa, ON, Canada, April
                  27-28, 2025},
  pages        = {156--167},
  publisher    = {{IEEE}},
  year         = {2025},
  url          = {https://doi.org/10.1109/Forge66646.2025.00026},
  doi          = {10.1109/FORGE66646.2025.00026},
  bibsource    = {dblp computer science bibliography, https://dblp.org}
}

@article{agent4se,
  author       = {Junwei Liu and
                  Kaixin Wang and
                  Yixuan Chen and
                  Xin Peng and
                  Zhenpeng Chen and
                  Lingming Zhang and
                  Yiling Lou},
  title        = {Large Language Model-Based Agents for Software Engineering: {A} Survey},
  journal      = {CoRR},
  volume       = {abs/2409.02977},
  year         = {2024},
  url          = {https://doi.org/10.48550/arXiv.2409.02977},
  doi          = {10.48550/ARXIV.2409.02977},
  eprinttype    = {arXiv},
  eprint       = {2409.02977},
  bibsource    = {dblp computer science bibliography, https://dblp.org}
}

@Inbook{AbdulKadir2024,
author="Abdul Kadir, Andi Fitriah
and Habibi Lashkari, Arash
and Daghmehchi Firoozjaei, Mahdi",
title="Android Operating System",
bookTitle="Understanding Cybersecurity on Smartphones: Challenges, Strategies, and Trends",
year="2024",
publisher="Springer Nature Switzerland",
address="Cham",
pages="25--42",
isbn="978-3-031-48865-8",
doi="10.1007/978-3-031-48865-8_2",
url="https://doi.org/10.1007/978-3-031-48865-8_2"
}

@misc{langchain_langgraph_2024,
author = {LangChain Inc.},
title = {LangGraph: Balance Agent Control with Agency},
year = {2024},
url = {https://www.langchain.com/langgraph},
howpublished = {Web page}
}

@misc{openai_gpt41_2025,
    author = {OpenAI},
    title = {Introducing GPT-4.1 in the API},
    year = {2025},
    month = {04},
    day = {14},
    url = {https://openai.com/index/gpt-4-1/},
    howpublished = {Web page},
}

@misc{anthropic_claude3.5_2024,
    author = {Anthropic},
    title = {Introducing computer use, a new Claude 3.5 Sonnet, and Claude 3.5 Haiku},
    year = {2024},
    month = {10},
    day = {22},
    url = {https://www.anthropic.com/news/3-5-models-and-computer-use},
    howpublished = {Web page},
}

@misc{anthropic_claude_sonnet_4_2025,
    author = {Anthropic},
    title = {Claude Sonnet 4},
    year = {2025},
    month = {05},
    day = {22},
    url = {https://www.anthropic.com/claude/sonnet},
    howpublished = {Web page},
}

@misc{qwen3lm_qwen3_coder_2024,
    author = {Alibaba Cloud (implied, as Qwen series is developed by Alibaba Cloud)},
    title = {Qwen3 Coder - Agentic Coding Adventure},
    year = {2024},
    url = {https://qwen3lm.com/},
    howpublished = {Web page},
}

@article{han2024survey,
  title={Survey: The Evolution and Future of Android Software Development},
  author={Han, Renda},
  journal={Deep Learning and Pattern Recognition},
  volume={1},
  number={1},
  year={2024}
}

@article{Kotlin-ML,
  author       = {Sergey Titov and
                  Mikhail Evtikhiev and
                  Anton Shapkin and
                  Oleg Smirnov and
                  Sergei Boytsov and
                  Dariia Karaeva and
                  Maksim Sheptyakov and
                  Mikhail Arkhipov and
                  Timofey Bryksin and
                  Egor Bogomolov},
  title        = {Kotlin {ML} Pack: Technical Report},
  journal      = {CoRR},
  volume       = {abs/2405.19250},
  year         = {2024},
  url          = {https://doi.org/10.48550/arXiv.2405.19250},
  doi          = {10.48550/ARXIV.2405.19250},
  eprinttype    = {arXiv},
  eprint       = {2405.19250},
  bibsource    = {dblp computer science bibliography, https://dblp.org}
}

@misc{aliyun_tongyi_llm_2024,
  author = {Alibaba Cloud},
  title = {Tongyi Large Language Models: The First Choice for Enterprises Embracing the AI Era},
  year = {2024},
  url = {https://www.aliyun.com/product/tongyi},
  howpublished = {Web page}
}

@article{LLM-Based_Domain_Modeling,
  author       = {Ru Chen and
                  Jingwei Shen and
                  Xiao He},
  title        = {A Model Is Not Built By {A} Single Prompt: LLM-Based Domain Modeling
                  With Question Decomposition},
  journal      = {CoRR},
  volume       = {abs/2410.09854},
  year         = {2024},
  url          = {https://doi.org/10.48550/arXiv.2410.09854},
  doi          = {10.48550/ARXIV.2410.09854},
  eprinttype    = {arXiv},
  eprint       = {2410.09854},
  bibsource    = {dblp computer science bibliography, https://dblp.org}
}

@incollection{robinson2024likert,
  title={Likert scale},
  author={Robinson, John},
  booktitle={Encyclopedia of quality of life and well-being research},
  pages={3917--3918},
  year={2024},
  publisher={Springer}
}

@inproceedings{goyal2007agile,
  title={Agile techniques for project management and software engineering},
  author={Goyal, Sadhna},
  booktitle={Major Seminar on Feature Driven Development},
  pages={1--19},
  year={2007}
}

@article{liualtdev,
  title={AltDev: Achieving Real-Time Alignment in Multi-Agent Software Development},
  author={Liu, Jie and Wang, Guohua and Yang, Ronghui and Zhao, Mengchen and Cai, Yi}
}

@article{AISD,
  author       = {Simiao Zhang and
                  Jiaping Wang and
                  Guoliang Dong and
                  Jun Sun and
                  Yueling Zhang and
                  Geguang Pu},
  title        = {Experimenting a New Programming Practice with LLMs},
  journal      = {CoRR},
  volume       = {abs/2401.01062},
  year         = {2024},
  url          = {https://doi.org/10.48550/arXiv.2401.01062},
  doi          = {10.48550/ARXIV.2401.01062},
  eprinttype    = {arXiv},
  eprint       = {2401.01062},
  bibsource    = {dblp computer science bibliography, https://dblp.org}
}

@inproceedings{evomac,
  author       = {Yue Hu and
                  Yuzhu Cai and
                  Yaxin Du and
                  Xinyu Zhu and
                  Xiangrui Liu and
                  Zijie Yu and
                  Yuchen Hou and
                  Shuo Tang and
                  Siheng Chen},
  title        = {Self-Evolving Multi-Agent Collaboration Networks for Software Development},
  booktitle    = {The Thirteenth International Conference on Learning Representations,
                  {ICLR} 2025, Singapore, April 24-28, 2025},
  publisher    = {OpenReview.net},
  year         = {2025},
  url          = {https://openreview.net/forum?id=4R71pdPBZp},
  bibsource    = {dblp computer science bibliography, https://dblp.org}
}

@inproceedings{whitehead2007collaboration,
  title={Collaboration in software engineering: A roadmap},
  author={Whitehead, Jim},
  booktitle={Future of Software Engineering (FOSE'07)},
  pages={214--225},
  year={2007},
  organization={IEEE}
}

@book{evans2004domain,
  title={Domain-driven design: tackling complexity in the heart of software},
  author={Evans, Eric},
  year={2004},
  publisher={Addison-Wesley Professional}
}

@incollection{bialy2017software,
  title={Software engineering for model-based development by domain experts},
  author={Bialy, M and Pantelic, V and Jaskolka, J and Schaap, A and Patcas, L and Lawford, M and Wassyng, A},
  booktitle={Handbook of system safety and security},
  pages={39--64},
  year={2017},
  publisher={Elsevier}
}

@inproceedings{petersen2009waterfall,
  title={The waterfall model in large-scale development},
  author={Petersen, Kai and Wohlin, Claes and Baca, Dejan},
  booktitle={International Conference on Product-Focused Software Process Improvement},
  pages={386--400},
  year={2009},
  organization={Springer}
}

@inproceedings{mandulapalli2025development,
  title={Development of Agentic Workflows with LangGraph for Software Development Life Cycle Automation},
  author={Mandulapalli, Shriraj and Hernandez, Emilio and Hall, Wayne Jordan and Chakeri, Alireza and Jaimes, Luis},
  booktitle={North American Conference on Industrial Engineering and Operations Management-Computer Science Tracks},
  pages={45--54},
  year={2025},
  organization={Springer}
}

@article{sharma2025optimised,
  title={Optimised Intelligent Software Company Management System using Multi-Agent Framework.},
  author={Sharma, Purva and Kaliappan, Jayakumar},
  journal={Grenze International Journal of Engineering \& Technology (GIJET)},
  volume={11},
  year={2025}
}

@inproceedings{akilesh2025multi,
  title={Multi-Agent hierarchical workflow for autonomous code generation with Large Language Models},
  author={Akilesh, S and Sekar, Rajeev and CU, Om Kumar and Prakalya, D and Suguna, M},
  booktitle={2025 IEEE International Students' Conference on Electrical, Electronics and Computer Science (SCEECS)},
  pages={1--5},
  year={2025},
  organization={IEEE}
}

@article{ge2023openagi,
  title={Openagi: When llm meets domain experts},
  author={Ge, Yingqiang and Hua, Wenyue and Mei, Kai and Tan, Juntao and Xu, Shuyuan and Li, Zelong and Zhang, Yongfeng and others},
  journal={Advances in Neural Information Processing Systems},
  volume={36},
  pages={5539--5568},
  year={2023}
}

@inproceedings{chen2023automated,
  title={Automated domain modeling with large language models: A comparative study},
  author={Chen, Kua and Yang, Yujing and Chen, Boqi and L{\'o}pez, Jos{\'e} Antonio Hern{\'a}ndez and Mussbacher, Gunter and Varr{\'o}, D{\'a}niel},
  booktitle={2023 ACM/IEEE 26th International Conference on Model Driven Engineering Languages and Systems (MODELS)},
  pages={162--172},
  year={2023},
  organization={IEEE}
}

@inproceedings{haryono2021androevolve,
  title={Androevolve: Automated update for android deprecated-api usages},
  author={Haryono, Stefanus A and Thung, Ferdian and Lo, David and Jiang, Lingxiao and Lawall, Julia and Kang, Hong Jin and Serrano, Lucas and Muller, Gilles},
  booktitle={2021 IEEE/ACM 43rd International Conference on Software Engineering: Companion Proceedings (ICSE-Companion)},
  pages={1--4},
  year={2021},
  organization={IEEE}
}

@article{DBLP:journals/corr/abs-2406-10018,
  author       = {Junwei Liu and
                  Yixuan Chen and
                  Mingwei Liu and
                  Xin Peng and
                  Yiling Lou},
  title        = {{STALL+:} Boosting LLM-based Repository-level Code Completion with
                  Static Analysis},
  journal      = {CoRR},
  volume       = {abs/2406.10018},
  year         = {2024},
  url          = {https://doi.org/10.48550/arXiv.2406.10018},
  doi          = {10.48550/ARXIV.2406.10018},
  eprinttype    = {arXiv},
  eprint       = {2406.10018},
  bibsource    = {dblp computer science bibliography, https://dblp.org}
}

@article{lin2025codereviewqa,
  title={CodeReviewQA: The Code Review Comprehension Assessment for Large Language Models},
  author={Lin, Hong Yi and Liu, Chunhua and Gao, Haoyu and Thongtanunam, Patanamon and Treude, Christoph},
  journal={arXiv preprint arXiv:2503.16167},
  year={2025}
}

@article{abedini2025leveraging,
  title={Leveraging Large Language Models for Classifying App Users' Feedback},
  author={Abedini, Yasaman and Heydarnoori, Abbas},
  journal={arXiv preprint arXiv:2507.08250},
  year={2025}
}

@inproceedings{sejfia2024toward,
  title={Toward improved deep learning-based vulnerability detection},
  author={Sejfia, Adriana and Das, Satyaki and Shafiq, Saad and Medvidovi{\'c}, Nenad},
  booktitle={Proceedings of the 46th IEEE/ACM international conference on software engineering},
  pages={1--12},
  year={2024}
}

@inproceedings{ciniselli2024generalizability,
  title={On the generalizability of deep learning-based code completion across programming language versions},
  author={Ciniselli, Matteo and Martin-Lopez, Alberto and Bavota, Gabriele},
  booktitle={Proceedings of the 32nd IEEE/ACM International Conference on Program Comprehension},
  pages={99--111},
  year={2024}
}

@article{baudry2024generative,
  title={Generative AI to generate test data generators},
  author={Baudry, Benoit and Etemadi, Khashayar and Fang, Sen and Gamage, Yogya and Liu, Yi and Liu, Yuxin and Monperrus, Martin and Ron, Javier and Silva, Andr{\'e} and Tiwari, Deepika},
  journal={IEEE Software},
  volume={41},
  number={6},
  pages={55--64},
  year={2024},
  publisher={IEEE}
}

@inproceedings{abdullin2025test,
  title={Test Wars: A Comparative Study of SBST, Symbolic Execution, and LLM-Based Approaches to Unit Test Generation},
  author={Abdullin, Azat and Derakhshanfar, Pouria and Panichella, Annibale},
  booktitle={2025 IEEE Conference on Software Testing, Verification and Validation (ICST)},
  pages={221--232},
  year={2025},
  organization={IEEE}
}

@article{tauhid2025explainability,
  title={Explainability as a Compliance Requirement: What Regulated Industries Need from AI Tools for Design Artifact Generation},
  author={Tauhid Ullah Shah, Syed and Hussein, Mohammad and Barcomb, Ann and Moshirpour, Mohammad},
  journal={arXiv e-prints},
  pages={arXiv--2507},
  year={2025}
}

@article{da2025llms,
  title={LLMs and Stack Overflow discussions: Reliability, impact, and challenges},
  author={Da Silva, Leuson and Samhi, Jordan and Khomh, Foutse},
  journal={Journal of Systems and Software},
  pages={112541},
  year={2025},
  publisher={Elsevier}
}

@inproceedings{lingma,
  title={Alibaba lingmaagent: Improving automated issue resolution via comprehensive repository exploration},
  author={Ma, Yingwei and Yang, Qingping and Cao, Rongyu and Li, Binhua and Huang, Fei and Li, Yongbin},
  booktitle={Proceedings of the 33rd ACM International Conference on the Foundations of Software Engineering},
  pages={238--249},
  year={2025}
}

@article{rahardja2025can,
  title={Can Agents Fix Agent Issues?},
  author={Rahardja, Alfin Wijaya and Liu, Junwei and Chen, Weitong and Chen, Zhenpeng and Lou, Yiling},
  journal={arXiv preprint arXiv:2505.20749},
  year={2025}
}

@article{luo2025rpg,
  title={RPG: A Repository Planning Graph for Unified and Scalable Codebase Generation},
  author={Luo, Jane and Zhang, Xin and Liu, Steven and Wu, Jie and Liu, Jianfeng and Huang, Yiming and Huang, Yangyu and Yin, Chengyu and Xin, Ying and Zhan, Yuefeng and others},
  journal={arXiv preprint arXiv:2509.16198},
  year={2025}
}

@article{peng2025code,
  title={Code Digital Twin: Empowering LLMs with Tacit Knowledge for Complex Software Development},
  author={Peng, Xin and Wang, Chong},
  journal={arXiv preprint arXiv:2503.07967},
  year={2025}
}

\end{document}